\documentclass[]{interact}

\usepackage{epstopdf}
\usepackage{amsmath,amssymb} 
\usepackage{graphicx} 
\usepackage{subcaption} 
\usepackage{hyperref}   
\usepackage{setspace}

\usepackage[numbers,sort&compress]{natbib}
\bibpunct[, ]{[}{]}{,}{n}{,}{,}

\theoremstyle{plain}

\theoremstyle{definition}

\theoremstyle{remark}

\begin{document}

\articletype{RESEARCH ARTICLE}

\title{Temporal and Content Coupling Analysis of Social Media User Behavior}

\author{
\name{Jipeng Tan\textsuperscript{a,b}, Mengye Yang\textsuperscript{a,b}, Zhanghao Li\textsuperscript{c} and Yong Min\textsuperscript{a,b,*}\thanks{* Corresponding author. Email address: myong@bnu.edu.cn}}
\affil{\textsuperscript{a}School of Journalism and Communication, Beijing Normal University, Beijing 100875, China; \textsuperscript{b}Computational Communication Research Center, Beijing Normal University, Zhuhai 519087, China; \textsuperscript{c}School of Journalism and Communication, Guangzhou University, Guangzhou, 510006, China}
}

\maketitle

\begin{abstract}
News consumption behavior is shaped by the coupling between temporal dynamics and content selection. This study proposes a multi-scale temporal-content framework and validates it on two large real-world news datasets, MIND and Adressa. Results reveal hierarchical temporal patterns. At the macroscale, Fourier modeling identifies clear circadian rhythms; at the mesoscale, session intervals follow a power-law distribution with $\alpha \approx 1$; and at the microscale, within-session action counts and inter-action intervals follow exponential distributions with $\lambda \approx 0.3$ and $\lambda \approx 0.02$, respectively. Content analysis shows that clicks are mainly driven by historical interests, while this dependence weakens as content diversity increases. Temporal-content coupling further indicates that users' historical interests dominate active time periods in shaping behavior. Preference groups also differ: timeliness and entertainment-oriented users click more frequently and rely more on historical interests, whereas diversified users click less and are more sensitive to content diversity.
\end{abstract}

\begin{keywords}
User Behavior; Temporal Dynamics; Content Selection Mechanisms; Temporal-Content Coupling
\end{keywords}

\section{Introduction}
\label{sec:introduction}

In the digital era, the widespread adoption of the internet has removed the spatiotemporal constraints on news consumption. The proliferation of smart devices has enabled users to access information beyond traditional temporal boundaries, resulting in complex temporal dynamics characterized by alternating periods of high-frequency, fragmented interactions and periodic deep reading~\cite{ref1,ref2}. Moreover, users' content selection is driven by a combination of historical interest dependence and exploratory diversity~\cite{ref3,ref4}. With growing attention to user behavior, AI agents driven by large language models (LLMs), such as ChatGPT and DeepSeek, have shown remarkable capabilities in understanding instructions, generating responses autonomously, and mimicking human-like behaviors~\cite{ref5,ref6,ref7}. However, whether these simulated behaviors truly reflect human cognitive and decision-making mechanisms is still an open question that requires empirical verification. Therefore, studying user behavior is not only essential for uncovering behavioral patterns, but also crucial for evaluating whether AI agents authentically model human cognitive processes.

Previous studies have contributed to our understanding of user news consumption behaviors, particularly in the modeling of user behavior~\cite{ref8,ref9}. Many of these studies have focused on analyzing users' consumption time patterns as one direction, while another stream has aimed at optimizing personalized recommendation systems based on the influence of historical interests on content selection~\cite{ref10,ref11,ref12,ref13,ref14}. Numerous studies have demonstrated that user behavior exhibits periodic patterns influenced not only by the content of the news but also by users' biological rhythms, daily routines, and social interactions~\cite{ref15}. For instance, Zhang et al.~\cite{ref16} analyzed large-scale mobile communication data and revealed that users' social connections follow a power-law distribution, exhibiting variations across age and temporal dimensions. Further research based on the WeChat platform demonstrated that highly active users maintain low volatility and high regularity in their posting behaviors, reflecting how platform usage habits shape behavioral rhythms~\cite{ref17}. Similarly, Kwon et al.~\cite{ref18}, in a study of Wikipedia, discovered that while individual editing behaviors conform to a power-law distribution, collective behaviors exhibit a dual power-law structure, highlighting the significant impact of collaborative interactions on overall behavior distribution. Although these studies primarily focus on social behaviors, the temporal regularities they reveal provide valuable insights into understanding news consumption. News consumption is not merely a random activity but is influenced by users' life rhythms, daily schedules, and fragmented time availability~\cite{ref19,ref20,ref21,ref22}.

In addition to temporal dynamics, users' historical interests significantly influence their click behaviors. Users exhibit a strong preference for content aligned with their historical interests, with click probabilities increasing as the semantic relevance of content rises~\cite{ref23,ref24,ref25,ref26,ref27,ref28}. Zhao et al.~\cite{ref29} identified a notable inertia effect in user interests, where users consistently focus on and periodically return to content reflecting their historical preferences. Fortunato et al.~\cite{ref30} further demonstrated that user search interests can increase the visibility of long-tail content, thereby mitigating the traffic bias inherent in search engines that favors popular pages, and enabling less popular, but more relevant, pages to receive clicks. However, this decision-making mechanism, which is based on historical interests, does not operate in isolation. On one hand, the exposure strategy of recommendation systems directly shapes the set of available options, requiring users to make selections from such content~\cite{ref31,ref32}. On the other hand, users' immediate decisions are influenced by the contextual characteristics of the presented content, including the presentation environment, the presence of simultaneously encountered content, and the diversity of topics and formats~\cite{ref33,ref34,ref35,ref36}. Therefore, understanding user click behaviors requires not only an analysis of their stable historical interests but also a quantification of the dynamic intervention effects of the exposure environment, which is critical for accurately modeling decision-making mechanisms in real-world scenarios.

Despite progress has been made in analyzing user news consumption behavior, particularly in the areas of temporal patterns and historical interest preferences, certain aspects remain to be addressed. First, most studies focus on short-term behaviors or single temporal scales, failing to systematically capture user behavior variations across multiple temporal scales, ranging from macroscopic to microscopic levels~\cite{ref37}. Second, while many studies emphasize the impact of historical interests on click behaviors, they often overlook the interaction between historical clicks and currently exposed content, which is crucial in user decision-making~\cite{ref38,ref39}. Finally, temporal dynamics and content selection are generally modeled separately, lacking a systematic framework that integrates their interactive evolution. Such a fragmented approach constrains the comprehensive understanding of user behaviors and limits the accurate modeling of their temporal-content evolution.

To overcome these limitations, this study proposes an innovative temporal-content coupling framework for user behavior analysis, systematically integrating temporal dynamics and content selection mechanisms. In the temporal dimension, it characterizes user behavior evolution across multiple temporal scales, ranging from the macroscopic 24-hour cycle to the mesoscale of session-level patterns and the microscopic level of single-action interactions. In the content dimension, it considers the combined influence of users' historical interest preferences and the diversity of currently exposed content on content selection. Finally, at the temporal-content coupling dimension, it further explores whether differences exist in temporal dynamics and content selection mechanisms among user groups with varying temporal activity patterns and content preferences. To validate the effectiveness of the proposed framework, this study conducts a systematic analysis of user behavior patterns at both individual and group levels using two large-scale, real-world news consumption datasets, MIND~\cite{ref40} and Adressa~\cite{ref41}. The framework reveals user behavior across temporal scales and content contexts, providing a solid foundation for understanding user behaviors. Furthermore, the proposed approach not only facilitates more accurate modeling of individual behaviors but also offers theoretical guidance for optimizing human-computer interaction design and modeling group behaviors. Ultimately, it establishes a foundational basis for advancing the reconstruction of the digital information ecosystem and promoting collaborative human-AI intelligence development.

\section{Data and Methods}
\label{sec:data_methods}

\subsection{Data Source Description}
To investigate user news consumption behavior, we employed two large-scale real-world datasets: MIND and Adressa. The MIND dataset, derived from Microsoft News, contains over 160,000 English articles and detailed exposure-click sequences from one million users between October 12 and November 22, 2019. The Adressa dataset, provided by Adresseavisen, includes longitudinal behavioral logs collected over ten weeks from January to March 2017. These datasets were selected for their complementary strengths. MIND offers rich content features and consistent click histories, making it ideal for modeling content selection mechanisms. Adressa, by contrast, provides high-resolution temporal traces, enabling fine-grained analysis of multi-scale temporal dynamics. Together, they form a robust empirical foundation for the proposed temporal-content coupling framework. For analysis, we retained only users with more than ten interactions to ensure data sufficiency. Moreover, for each user in the MIND dataset, a real-time historical click sequence for each exposure instance is constructed by appending the most recent click to the historical sequence from the previous exposure, thereby capturing the user's click behavior leading up to each decision.

\subsection{Methods}
\begin{figure}[t]
	\centering
	\includegraphics[width=0.9\textwidth]{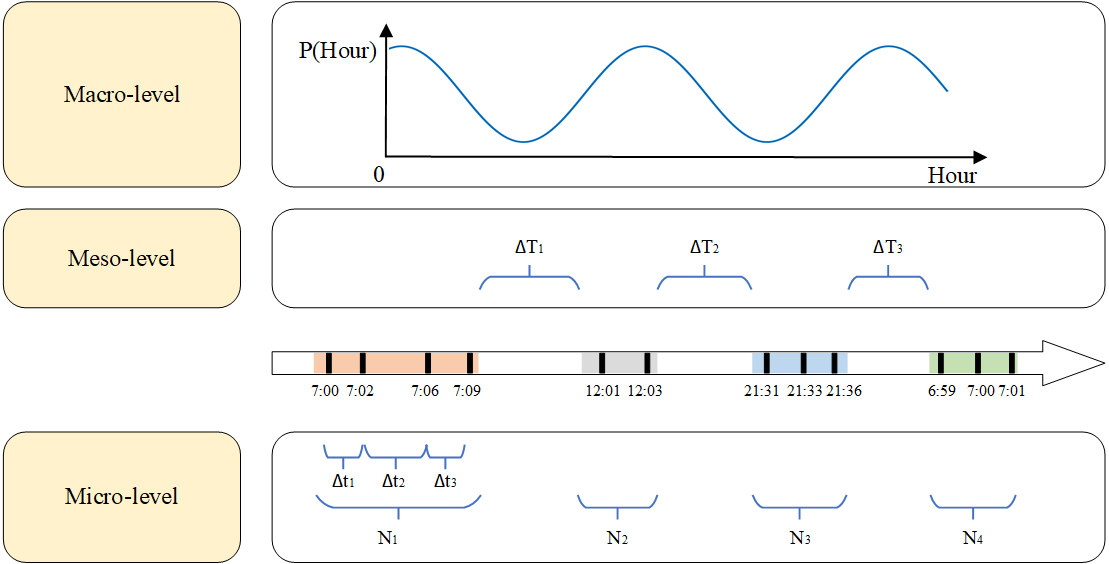}
	\caption{Schematic diagram of the dynamic time window division for user news consumption behavior.}
	\label{fig:1}
\end{figure}

To effectively capture user behavior patterns in the context of news consumption, this study proposes a temporal-content coupling user analysis framework. In the temporal dimension, a three-scale analysis framework is constructed, as shown in Figure~\ref{fig:1}. At the macroscopic level, the distribution of user click frequencies over a 24-hour period is statistically represented as $X(t) = \{x_0, x_1, \dots, x_{23}\},$where $X(t)$ denotes the proportion of news clicks at time $t$ relative to the total clicks for the day. Periodic features are extracted using a Fourier series,$X(t) = \sum_n \left( a_n \sin \frac{2 \pi n t}{T} \right. + \\
\left. b_n \cos \frac{2 \pi n t}{T} \right)$, modeling the daily rhythm of user news visits and high-frequency active periods. At the mesoscale level, user behavior sequences are divided by sessions, and the time interval between adjacent sessions is defined as $\Delta T$. Distribution modeling is used to reveal the regularity of users' news consumption rhythms. To improve the behavioral validity of session segmentation, a time-interval hierarchical clustering method is introduced, with a threshold $\Delta T_\tau$ set to dynamically adapt to the operational behaviors of different users, preventing the inclusion of short breaks during news browsing (e.g., switching applications, checking messages, etc.). At the microscopic level, the intra-session operational intervals ($\Delta t$) and the number of actions ($N$) are further extracted to model short-term reading rhythms. This multi-scale modeling framework not only enhances the expressiveness of temporal features but also provides more interpretable structural support for characterizing the dynamics of user interests.

\begin{figure}[t]
	\centering
	\includegraphics[width=0.9\textwidth]{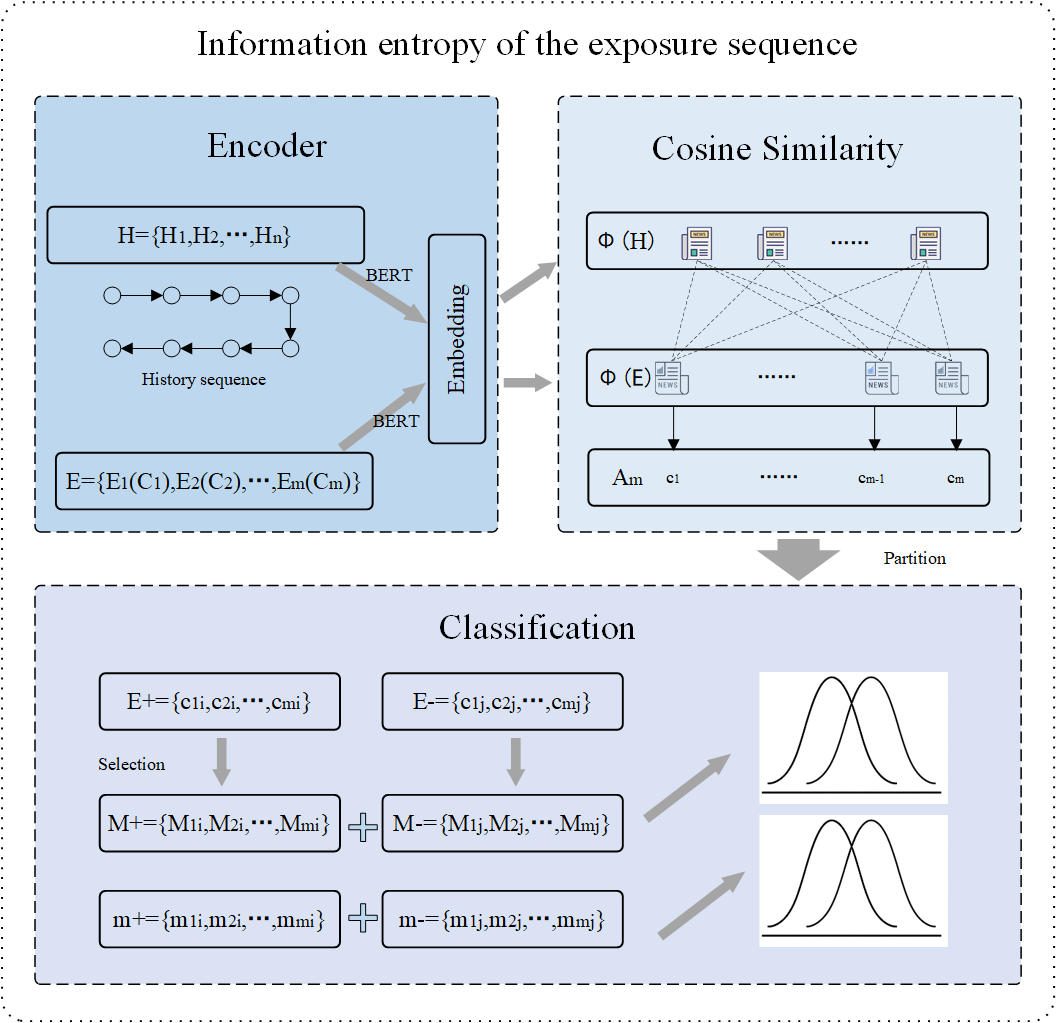}
	\caption{The schematic diagram of the content analysis process for user news consumption behavior}
	\label{fig:2}
\end{figure}

In the content dimension, to investigate the influence of historical interests and the current exposure sequence on user click behavior, this study proposes a modeling framework that integrates semantic similarity and content structural features, with a single exposure as the unit of analysis. As shown in Figure~\ref{fig:2}, each exposure is represented as $(H, E)$, where $H = \{h_1, h_2, \dots, h_n\}$ is the historical click sequence, and $E = \{e_1, e_2, \dots, e_m\}$ is the current exposure sequence. For each news item in the exposure sequence, a click label $C$ is assigned, where $C = \{0,1\}$, with 0 indicating no click and 1 indicating a click on the exposure. For the title content of both $H$ and $E$, we employ the BERT model for semantic encoding. We obtain embedding vectors $\phi(H)$ for the historical click sequence and $\phi(E)$ for the exposure sequence. For each exposure $e_i$ and historical click $h_j$, the cosine similarity is computed as:
\begin{equation}
	\mathrm{sim}(e_i, h_j) = \frac{\phi(e_i)^T \phi(h_j)}{\|\phi(e_i)\| \|\phi(h_j)\|}.
	\label{eq:cosine_similarity}
\end{equation}
\\
This results in a similarity matrix $A_{m \times n} = \{c_1, c_2, \dots, c_m\}^T$, where $c_i = \{s_{i1}, s_{i2}, \dots, s_{in}\}$ represents the similarity sequence between the $i$th exposure title and the historical click titles. Here, $s_{ij}$ denotes the similarity between the $i$th exposure title and the $j$th historical click. To further analyze the impact of historical interests on click behavior, we divide the exposure samples into the exposure-click set $E^{+} = \{e | C=1\}$ and the exposure-no-click set $E^{-} = \{e | C=0\}$ based on the click label $C$. For each exposure, we extract the maximum value ($M$) and the median ($m$) from the corresponding similarity sequence $c_i$ as content structural features. The maximum value represents the strength of association between the exposure content and the most similar historical interest, with a higher value indicating a stronger user response to the most similar historical content. The median reflects the overall matching level of user interests within a broader interest space, with a higher median suggesting more stable and consistent interest in the exposure content~\cite{ref42}. By comparing the distribution differences of the maximum value and median features between the sets $E^{+}$ and $E^{-}$, we quantify the role of historical interests in the current exposure click decision-making process.

To further investigate the impact of exposure content on user click behavior, this study introduces two metrics, information entropy and Euclidean distance, to model the diversity of the exposure sequence. The information entropy of the exposure sequence is defined as: $\mathrm{En}(E) = - \sum_{i=1}^n p(x_i) \log p(x_i)$, where $p(x_i)$ represents the probability distribution of the category of the news item in the current exposure sequence. A higher information entropy indicates greater overall diversity. Additionally, based on the embedding vector $\phi(E)$, the Euclidean distance between each pair of news items in the exposure sequence is calculated, and the mean value of these distances is taken as another measure of the diversity of the exposure sequence.

Under different levels of exposure sequence diversity, this study analyzes the changes in user click behavior from two perspectives. First, it examines the distributional changes of the click samples at different levels of exposure diversity to assess the impact of content diversity on the user selection mechanism. Second, since the number of clicked samples is significantly smaller than that of non-clicked samples, the non-clicked samples are used to approximate the overall trend of the clicked samples. It then further computes the Wasserstein (WS) distance between the distributions of clicked and non-clicked samples. A larger WS distance indicates that the minimum transportation cost between the two distributions is greater, meaning that the transition from the exposure distribution to the click distribution is more difficult. The greater the degree of influence from historical interests on clicks, the more challenging the transition. Through a comparative analysis of these dual metrics, the study systematically reveals the mechanism by which exposure content diversity shapes the user selection process.

Finally, this study conducts a coupled analysis of user news consumption behavior from two dimensions: time-oriented and content-oriented. In the temporal dimension, clustering methods such as K-means and Gaussian Mixture Models (GMM) are applied to the user's 24-hour active time distribution feature $X(t)$ to identify user groups during different active periods. In the content dimension, user click frequencies on news categories are statistically analyzed to extract the top-$n$ interest tags. After encoding, various clustering algorithms such as K-means, GMM, Hierarchical Clustering, and Birch are employed for grouping. The clustering effectiveness is evaluated using silhouette scores, and the best clustering model and parameter settings are selected accordingly. Based on the clustering results, this study further investigates the differences and commonalities in time behavior patterns and content selection mechanisms between user groups at different active stages and those with different interest preferences, providing a systematic perspective for a deeper understanding of the dynamic characteristics and multi-scale coupling mechanisms of user news consumption behavior.

\section{Results}
\label{sec:results}

Based on the previously proposed temporal-content coupling analysis framework, this section explores the user behavior patterns in the MIND and Adressa datasets. To comprehensively validate the analytical capabilities of the framework, we not only examine the behavioral characteristics at the group level but also select typical individual cases for in-depth analysis.

\subsection{Analysis of Temporal Dimensions}
\label{subsec:temporal_analysis}

\begin{figure}[t]
\centering
\begin{minipage}[t]{0.48\linewidth}
  \centering
  \includegraphics[height=5cm,width=\linewidth,keepaspectratio]{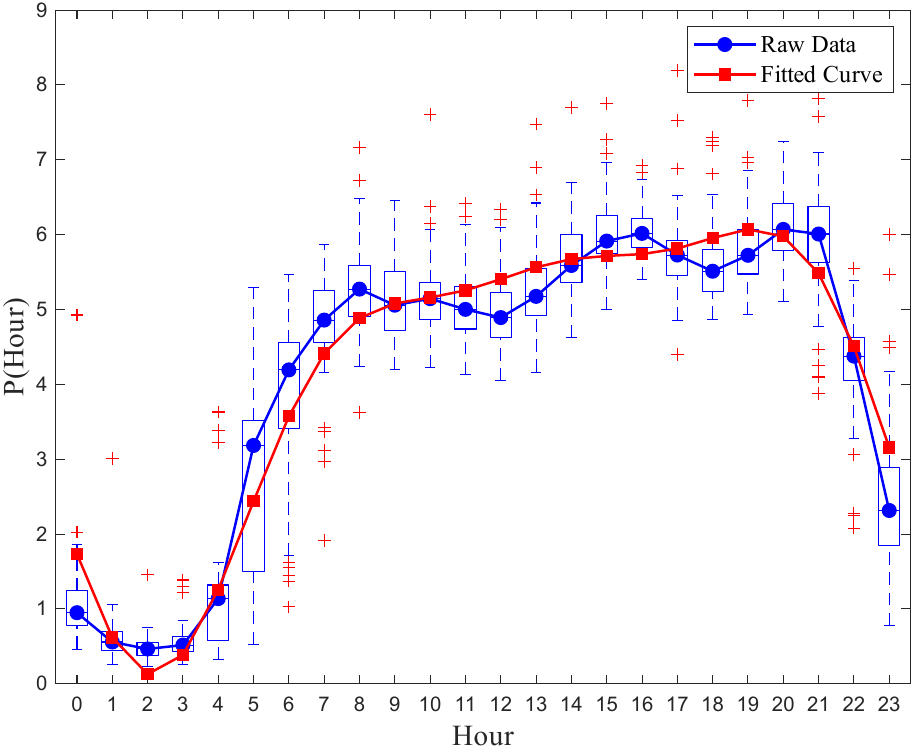}\\
  \small (a)
\end{minipage}%
\hspace{2pt}%
\begin{minipage}[t]{0.48\linewidth}
  \centering
  \includegraphics[height=5cm,width=\linewidth,keepaspectratio]{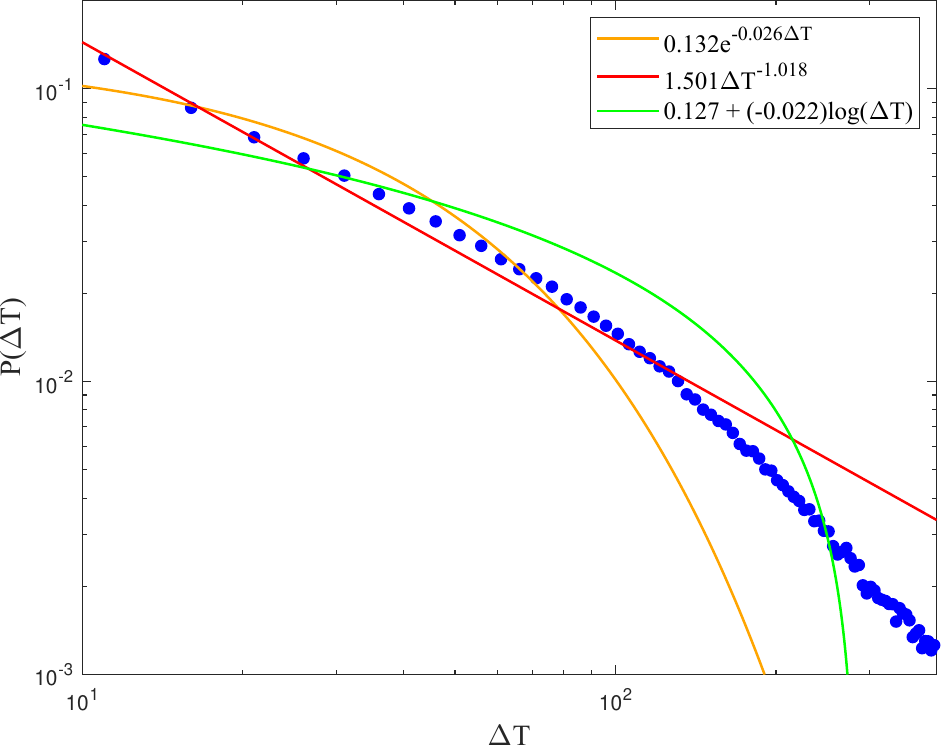}\\
  \small (b)
\end{minipage}

\vspace{1pt}
\begin{minipage}[t]{0.48\linewidth}
  \centering
  \includegraphics[height=5cm,width=\linewidth,keepaspectratio]{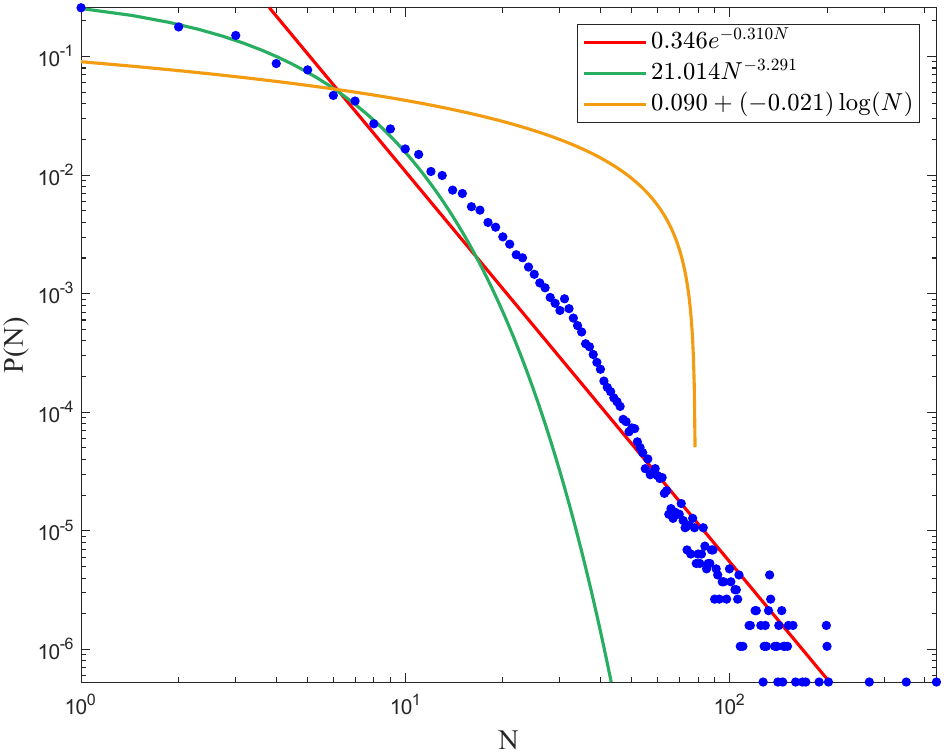}\\
  \small (c)
\end{minipage}%
\hspace{2pt}%
\begin{minipage}[t]{0.48\linewidth}
  \centering
  \includegraphics[height=5cm,width=\linewidth,keepaspectratio]{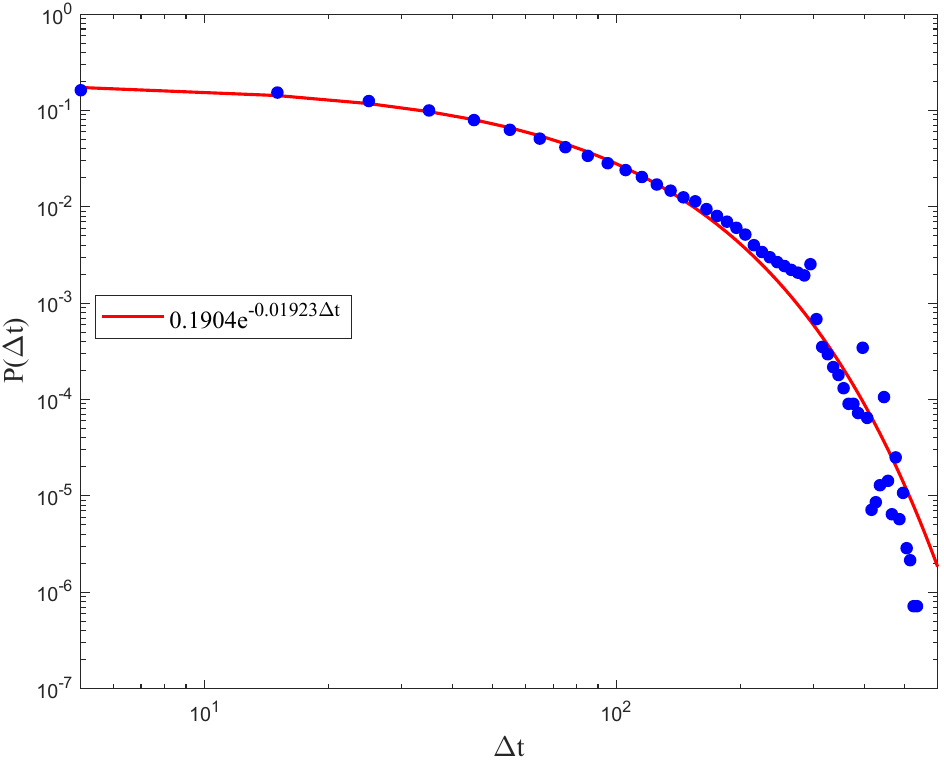}\\
  \small (d)
\end{minipage}

\caption{
    (a) Analysis and fitting comparison of the temporal pattern of news consumption activity frequency at the macroscopic dimension; 
    (b) Distribution of session intervals and fitting function under $\Delta T < \Delta T'$ at the mesoscopic dimension; 
    (c) Function of $P(N)$ with respect to $N$ at the microscopic dimension;
    (d) Function of $P(\Delta t)$ with respect to $\Delta t$ at the microscopic dimension.
} 
\label{fig:3}
\end{figure}

To systematically investigate the multi-scale temporal patterns of user behavior, the threshold in this study is set to $\Delta T_{\tau} = 10$ min, referencing the work of He and Göker~\cite{ref43}. Each session is categorized into the corresponding hourly period based on the session's starting timestamp, and the proportion of sessions in each period relative to the total number of sessions on that day is calculated. A box plot is used to display the distribution of 24-hour period data (Figure.~\ref{fig:3}(a)). To further quantify the periodic regularity, a Fourier series model is constructed, where $T = 24$ hours is the fundamental period and $k=3$ is the number of harmonics. Goodness-of-fit tests show that the model has a high explanatory power for the data ($R^2 = 0.957$, $p = 4.35 \times 10^{-13}$). To validate the generalizability of the model, seven randomly selected days of data from the dataset were used, and the fitted results are consistent with the data (as shown in Supplementary Figure A1), further confirming the effectiveness and robustness of the Fourier series in characterizing the periodic patterns of user behavior. This result suggests that user news consumption behaviors exhibit a stable circadian rhythm, closely aligning with human biological clock-driven activity patterns. The fitted Fourier series function is expressed as:
\begin{equation}
	\begin{split}
		f(T) =\; &4.1667 + (-1.7887) \cos\left(\frac{2\pi T}{24}\right) + (-1.6089) \sin\left(\frac{2\pi T}{24}\right) \\
		&+ (-0.5976) \cos\left(\frac{4\pi T}{24}\right) + (-1.1159) \sin\left(\frac{4\pi T}{24}\right) \\
		&+ (-0.0467) \cos\left(\frac{6\pi T}{24}\right) + (-0.4184) \sin\left(\frac{6\pi T}{24}\right).
	\end{split}
	\label{eq:fourier_series}
\end{equation}

\begin{figure}[t]
	\centering
	\includegraphics[width=0.6\textwidth]{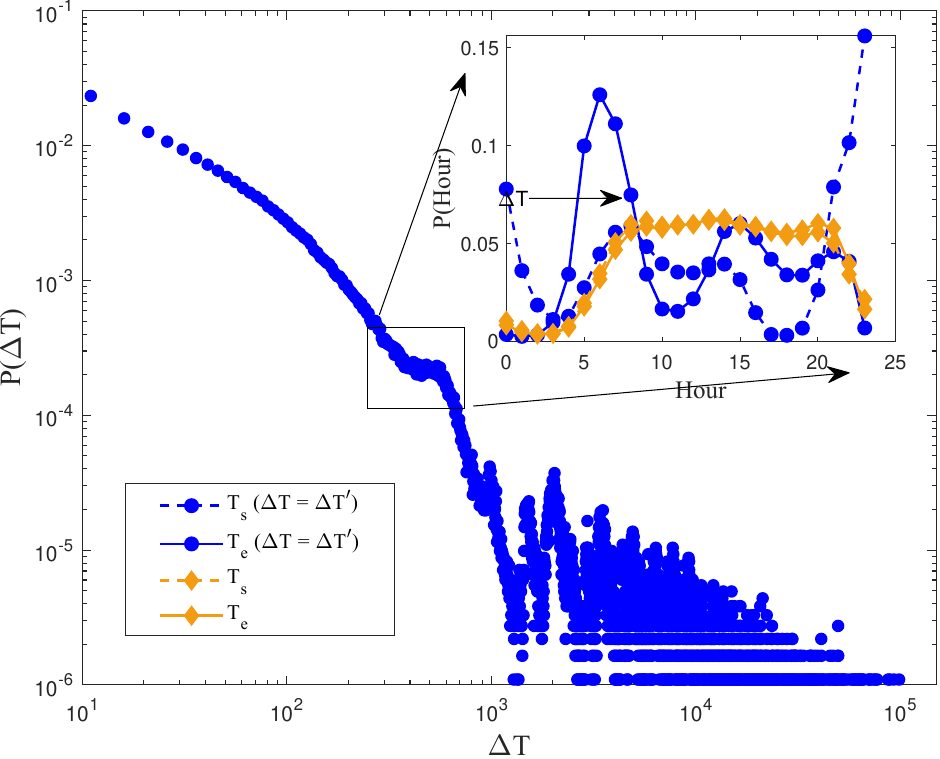}
	\caption{The probability density function distribution of time intervals between sessions. The subfigures show the frequency distribution of interval start and end times relative to the 24-hour cycle under $\Delta T = \Delta T'$ and $\Delta T < \Delta T'$.}
	\label{fig:4}
\end{figure}

After investigating the macroscopic temporal periodicity, this study analyzes user behavior at the mesoscale and microscale. At the mesoscale, the focus is on the distribution characteristics of session intervals $\Delta T$ (Figure~\ref{fig:4}). The results show that the probability density function $P(\Delta T)$ decreases with increasing $\Delta T$, with a distinct inflection point $\Delta T'$. Moreover, $P(\Delta T)$ exhibits a significant bimodal distribution. When $\Delta T < \Delta T'$, short interval behaviors may follow a power-law distribution, indicating that users exhibit high-frequency fragmented access patterns, which align with the widely observed self-organized criticality in social behaviors. For example, users may quickly access news during commuting or break times, which accounts for approximately 80\% of their daily activity, thereby establishing stable daily consumption patterns. When $\Delta T = \Delta T'$ (with $360\,\mathrm{min} \leq \Delta T' \leq 540\,\mathrm{min}$), we select matching time pairs, each consisting of two timestamps $T_s$ and $T_e$, representing the earliest and latest times of the interval start. In the subfigure of Figure~\ref{fig:4}, the distributions of $T_s$ and $T_e$ at different $\Delta T$ are plotted. The results show that, when $\Delta T$ is not distinguished, the distributions of $T_s$ and $T_e$ are consistent with the macroscopic consumption behavior shown in Figure~\ref{fig:3}(a). When $\Delta T = \Delta T'$, the distributions of $T_s$ and $T_e$ are concentrated in the nighttime period, indicating that user access is interrupted by sleep behaviors and resumes after waking up, creating a natural break in the behavioral pattern. Therefore, it is necessary to exclude long intervals $\Delta T \geq \Delta T'$ when simulating user access behavior. Figure~\ref{fig:3}(b) shows the distribution of $P(\Delta T)$ and $\Delta T$ when $\Delta T < \Delta T'$, along with fittings using power-law, exponential, and logarithmic functions. We find that the power-law function performs the best. The fitting equation for $P(\Delta T)$ is: $P(\Delta T) = 1.501 \times \Delta T^{-1.018}$.

Next, we focus on the microscopic features within sessions by extracting the number of actions $N$ in each session and the time difference $\Delta t$ between adjacent actions. Figure~\ref{fig:3}(c) shows the distribution of $N$ on a double-logarithmic scale. It can be observed that the distribution of action counts is mainly concentrated between 1 and 8 actions, with the high-frequency region located at the lower end of the action count. Therefore, it is crucial to emphasize the probability distribution at the lower action counts. We find the frequency exhibits an exponential decay as the number of actions increases. The specific fitting formula is: $P(N) = 0.346 e^{-0.310 N}$. Figure~\ref{fig:3}(d) shows the distribution of $\Delta t$ on a double-logarithmic scale. From Figure~\ref{fig:3}(d), it is evident that $\Delta t$ also follows an exponential decay pattern, with the high-frequency region concentrated between 0 and 100 seconds, and the frequency approaching zero beyond 200 seconds. These patterns reflect the users' high-frequency interactions in short periods, exhibiting behaviors aimed at quickly obtaining information, while larger time differences may correspond to short pauses or departures. The specific fitting formula is: $P(\Delta t) = 0.190 e^{-0.019 \Delta t}$. Similar distribution patterns for the triple time distribution rules in the MIND dataset are also observed (see Supplementary Figure A2).

\begin{figure}[htbp]
	\centering
	\includegraphics[width=0.6\textwidth]{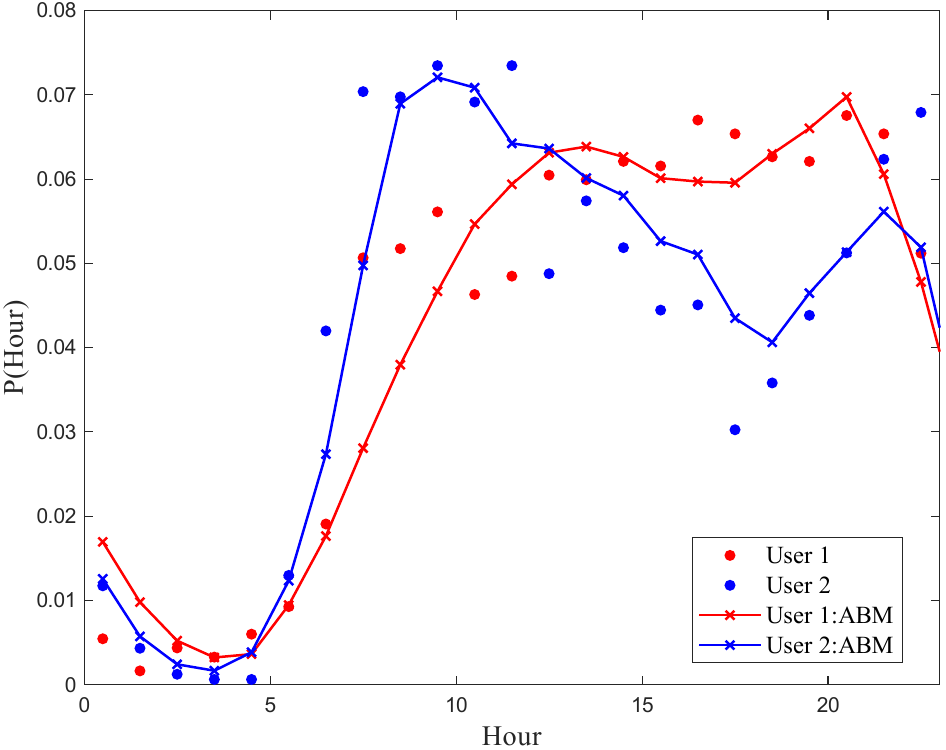}
	\caption{ABM simulation of user activity compared to real user activity over a 24-hour period.}
	\label{fig:5}
	
\end{figure}

After analyzing the temporal behavior patterns at the group level, this study further selected two users randomly from the Adressa dataset for a detailed examination of their individual behavior patterns. We found that these two users exhibit behavior characteristics at both the mesoscale and microscale that align with those observed in the overall group. Specifically, at the mesoscale, the session intervals follow a power-law distribution, while at the microscale, both the number of actions within a session and the inter-action intervals between actions exhibit an exponential distribution (see Supplementary Figure A3). Additionally, based on the triple temporal scale framework, the study introduces an agent-based modeling approach to simulate users' click behavior trajectories within a 24-hour cycle, aiming to capture the temporal characteristics of user clicks. Specifically, the first click time for each agent is determined by probability sampling from $X(t)$, and subsequent click times are determined by the empirical distribution of the time interval $\Delta T$, represented as: $T_{t+1} = T_t + \Delta T$. Furthermore, the model introduces a dynamic exit probability based on different time periods, meaning that after each click, the probability of continuing to generate the next behavior is determined. The exit probabilities are categorized into four types based on time periods, as follows:
\begin{equation}
	\begin{split}
		p = p_i, \quad i \in \{\mathrm{MP}, \mathrm{DP}, \mathrm{EP}, \mathrm{LNP}\},
	\end{split}
	\label{eq:ABM}
\end{equation}\\
where $p_{\mathrm{MP}}$, $p_{\mathrm{DP}}$, $p_{\mathrm{EP}}$, and $p_{\mathrm{LNP}}$ represent the exit probabilities for the morning peak, daytime, evening, and late-night periods, respectively. The exit probability is designed to reflect the user’s behavior rhythm in real life. Additionally, to prevent excessively high-frequency clicking, we impose a restriction that the time interval between each click must exceed $\Delta T_\tau$, thereby avoiding implausibly frequent click sequences. If the next click time exceeds one day, it is processed according to the periodicity rule. At the session level, we use the click count and click interval distribution from the microscale to model the operational time for micro-level actions. Figure~\ref{fig:5} shows the activity levels of different agents and real users over 24 hours. It can be observed that the proposed analytical framework can, to some extent, simulate the click temporal dynamics of real users.

\subsection{Analysis of Content Dimensions}

\begin{figure}[t]
	\centering
	\includegraphics[width=0.7\textwidth]{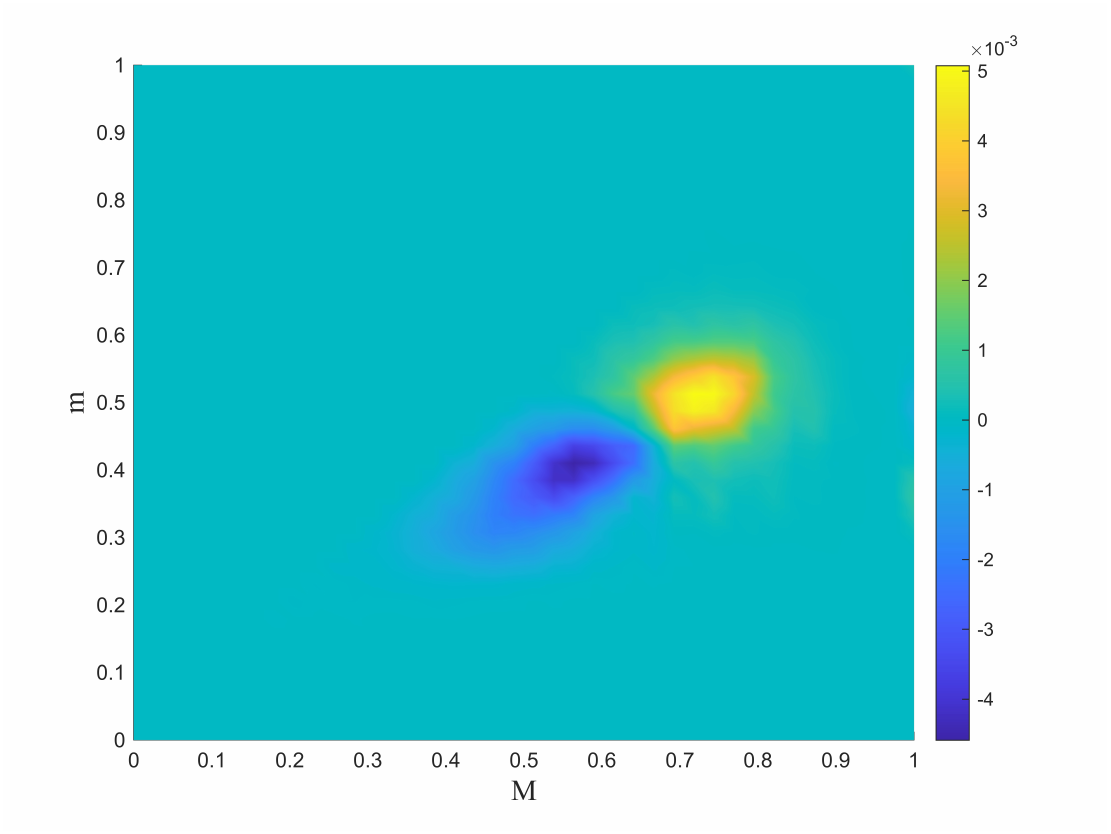}
	\caption{The difference between the joint probability density distributions of $E_m^{+}$ and $E_M^{+}$ and the joint probability density distributions of $E_m^{-}$ and $E_M^{-}$.}
	\label{fig:6}
	
\end{figure}

Based on the content dimension analysis framework, this study compares the semantic similarity distributions of the exposure-click set $E^{+}$ and the exposure-no-click set $E^{-}$ (Figure~\ref{fig:6}), revealing the historical dependence of user click behaviors. Figure~\ref{fig:6} displays the difference in the joint probability density distributions between $E_m^{+}$ and $E_M^{+}$ and between $E_m^{-}$ and $E_M^{-}$. The results show that, compared to $E^{-}$, the distribution of $E^{+}$ is significantly concentrated in the high-similarity regions of both the maximum value and the median, indicating that users are more likely to click on and tend to select news items with a higher similarity to their historical interests. Furthermore, this click behavior is driven not only by local similarities but is more reliant on the overall similarity between the content and historical interests. To address the issue of the absence of exposure logs in the Adressa dataset, this study introduces the proxy exposure hypothesis: treating the user's last click as an exposure click event $E^{+}$ and calculating its similarity to the historical click sequence. The results reveal patterns similar to those observed in the MIND dataset, as shown in Supplementary Figure A4.

\begin{figure}
\centering
\begin{minipage}[t]{0.48\linewidth}
  \centering
  \includegraphics[height=5cm, width=\linewidth, keepaspectratio]{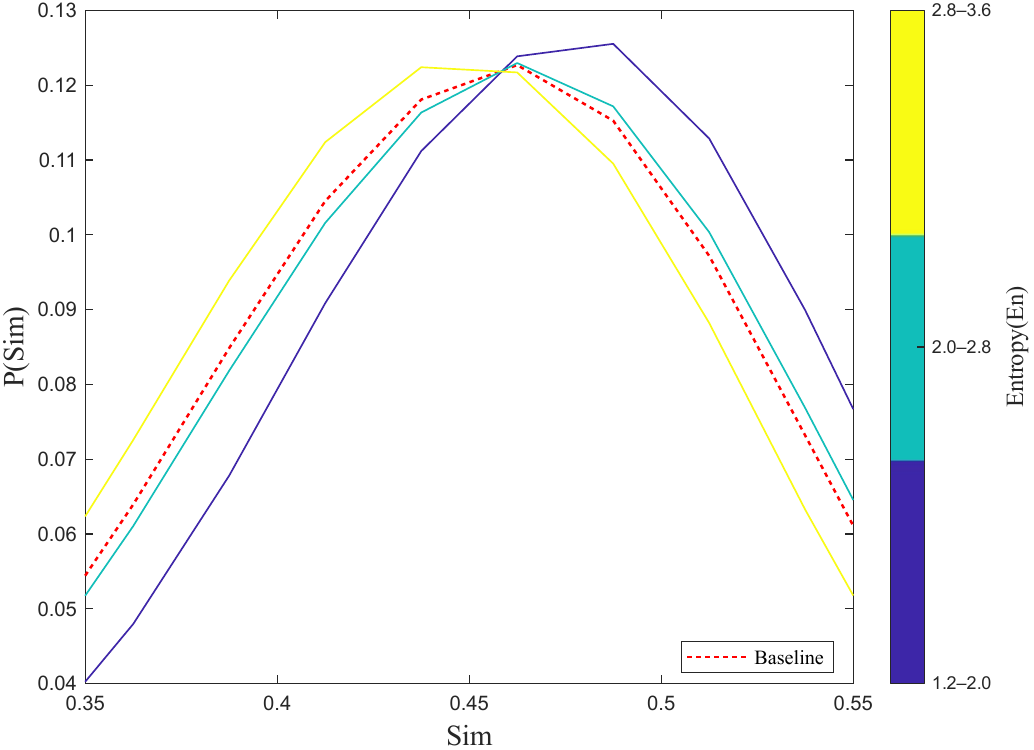}\\
  \small (a)
\end{minipage}%
\hspace{2pt}%
\begin{minipage}[t]{0.48\linewidth}
  \centering
  \includegraphics[height=5cm, width=\linewidth, keepaspectratio]{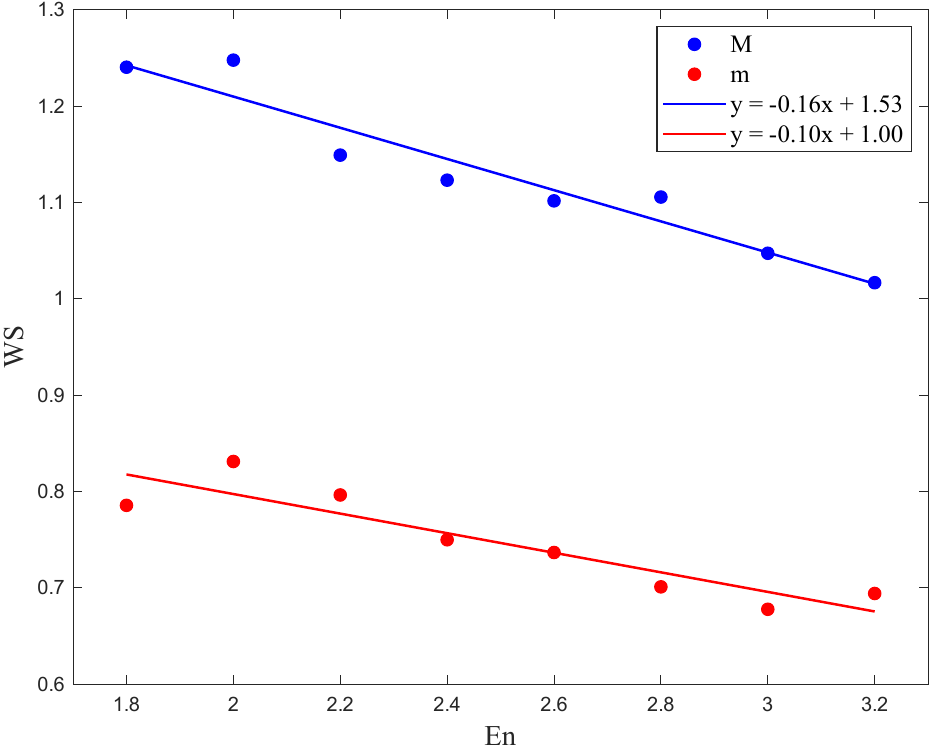}\\
  \small (b)
\end{minipage}
\caption{
  (a) Probability density distributions of $E_m^{+}$ and $E_m^{-}$; 
  (b) Function of $W(E_m^{+}, E_m^{-})$ and $W(E_M^{+}, E_M^{-})$ with respect to $\mathrm{En}$.
}
\label{fig:7}
\end{figure}

Next, we explore whether exposure diversity affects the degree of association between user clicks and content similar to their historical interests. To eliminate the interference of exposure content length on exposure diversity analysis, this study first performed data cleaning on the MIND dataset and filtered samples with exposure sequence lengths ranging from 10 to 15 for analysis. Based on the exposure diversity indicator $\mathrm{En}(E)$, which mainly concentrates in the range of 1.2 to 3.6 (details can be found in Supplementary Figure A5), this study divides the samples into three groups based on $\mathrm{En}(E)$: low diversity group (1.2–2.0), medium diversity group (2.0–2.8), and high diversity group (2.8–3.6). Figure~\ref{fig:7}(a) presents the probability density distribution of the semantic similarity between the exposure sequence and the historical sequence under varying exposure diversity conditions. The results show that, with an increase in exposure diversity, the overall similarity distribution shifts to the left. This indicates that the increase in exposure diversity weakens the influence of historical interests on user click behaviors. To further validate the robustness of the diversity metric, this study employs Euclidean distance as an alternative indicator for testing (see Supplementary Figure A6), and the results are consistent with the above conclusion, further reinforcing the reliability of the findings. Subsequently, we examine the distributional differences between the click distribution and the exposure list distribution under varying exposure diversity conditions. Figure~\ref{fig:7}(b) shows the WS distance as a function of $\mathrm{En}(E)$. The results indicate that the maximum value and the median exhibit a monotonically decreasing trend. This suggests that as exposure diversity increases, the degree to which user click behavior deviates from the exposure content gradually decreases, and user behavior begins to align more closely with the content currently presented to them, rather than being influenced by historical interests. Overall, exposure content diversity can reduce the impact of historical interests to some extent, making the user's clicked content more closely related to the currently exposed content.

To verify that users are more likely to select news articles with high similarity to their historical interests (i.e., the right-skewed phenomenon) and that increased exposure diversity weakens the influence of historical interests on click behavior, we designed a random experiment aimed at excluding potential system-induced effects and focusing on whether these phenomena are driven by user behavior. To achieve this, we introduced a social bot to perform random clicks on the Google News platform. Through this experiment, we were able to clearly distinguish between the effects of user behavior and the recommendation system mechanisms. The results show that these two phenomena are primarily driven by user behavior rather than the recommendation system (the detailed experimental design and results are discussed in the Random Click Behavior Experiment section in the supplementary materials).

\begin{figure}
\centering
\begin{minipage}[t]{0.48\linewidth}
  \centering
  \includegraphics[height=5cm, width=\linewidth, keepaspectratio]{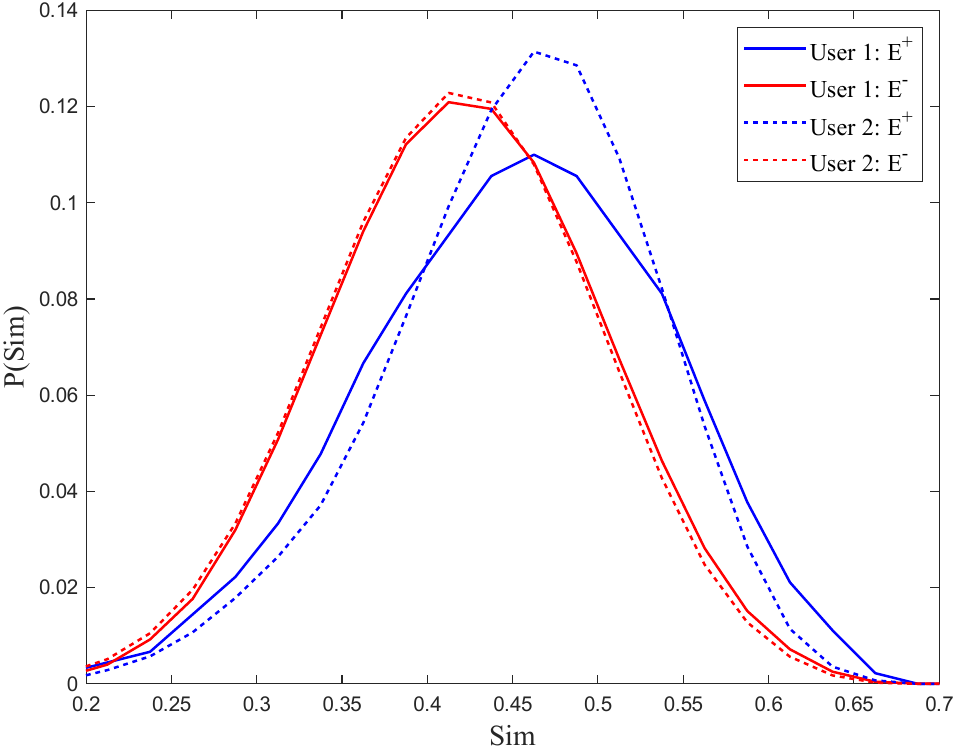}\\
  \small (a)
\end{minipage}%
\hspace{2pt}%
\begin{minipage}[t]{0.48\linewidth}
  \centering
  \includegraphics[height=5cm, width=\linewidth, keepaspectratio]{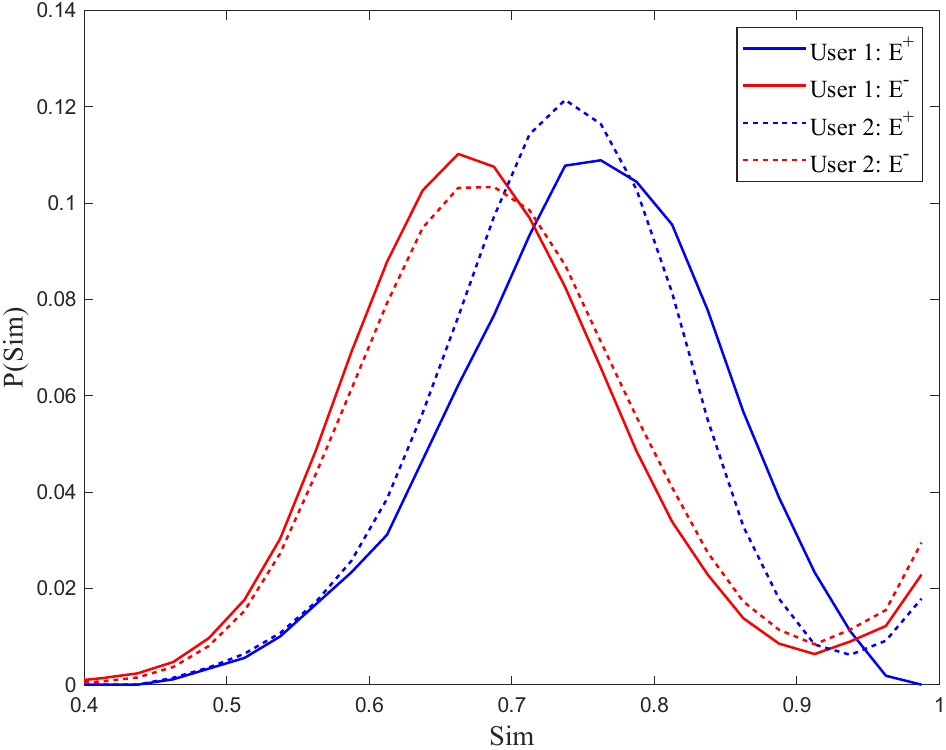}\\
  \small (b)
\end{minipage}
\caption{
    Randomly selected distributions of (a) $E_m^{+}$, $E_m^{-}$ and (b) $E_M^{+}$, $E_M^{-}$ for two users from the MIND dataset.
}
\label{fig:8}
\end{figure}

Similar to the analysis in the temporal dimension, we selected two users from the MIND dataset for analysis. Figure~\ref{fig:8}(a) shows the distribution of $E_m^{+}$ and $E_m^{-}$ for the two users, while Figure~\ref{fig:8}(b) presents the distribution of $E_M^{+}$ and $E_M^{-}$. At the individual level, we observe a pattern consistent with the group-level findings: for each user, the distribution of semantic similarity between clicked news items and their historical interests is generally significantly higher than that of non-clicked items. However, there is some variability across users in these similarity measures, indicating that while users overall tend to click on content similar to their historical interests, there exists a certain degree of individual heterogeneity.

\subsection{Integrated Temporal and Dimensional Analysis}

In the previous section, we systematically revealed the temporal patterns of user news consumption and the characteristics of content clicks, uncovering how user activity patterns at different time levels and their click preferences are influenced by historical data and exposure sequences. In this section, we examine the behavioral differences among users of various categories under two coupling relationships: time orientation and content orientation. In the content-oriented coupling analysis, based on the frequency distribution of user clicks across news categories, we extracted each user's top-3 dominant interest tags (details can be found in Supplementary Figure A7). Using the K-means clustering algorithm (with silhouette coefficients of 0.6341 for K-means, 0.5966 for GMM, 0.6214 for Hierarchical Clustering, and 0.5598 for Birch), we identified user groups with different content preferences, such as Timeliness Orientation (users with a preference for news content), In-Depth Content Orientation (users who favor paid content), Entertainment Orientation (users primarily interested in sports), and Diverse Interest Orientation (users exhibiting a broad range of interests). It is worth noting that, unlike Adressa, the MIND dataset shows a life-oriented preference group after clustering but does not contain users with an In-Depth Content Orientation. In the time-oriented clustering analysis, we adopted the K-means algorithm to cluster users’ 24-hour activity patterns, dividing users into daytime-active and nighttime-active groups. Through this dual-dimensional analysis framework, we compared the differences in temporal behavior characteristics and content selection mechanisms across different groups, further revealing the coupling relationship between time patterns and content preferences in user news consumption behavior.

\begin{figure}
\centering
\begin{minipage}[t]{0.48\linewidth}
  \centering
  \includegraphics[height=5cm, width=\linewidth, keepaspectratio]{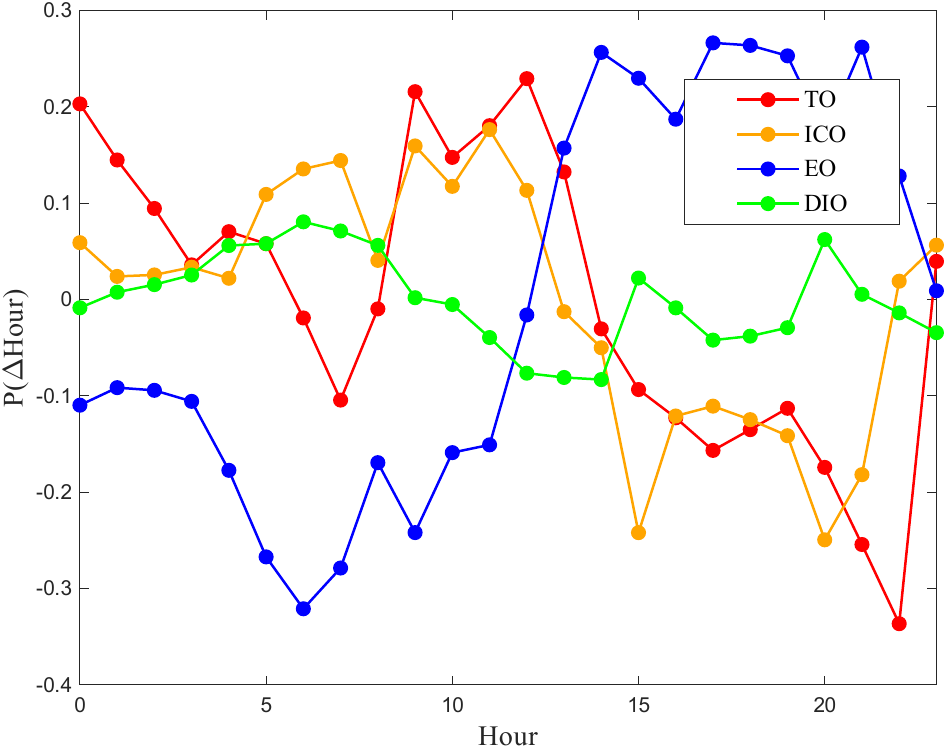}\\
  \small (a)
\end{minipage}%
\hspace{2pt}%
\begin{minipage}[t]{0.48\linewidth}
  \centering
  \includegraphics[height=5cm, width=\linewidth, keepaspectratio]{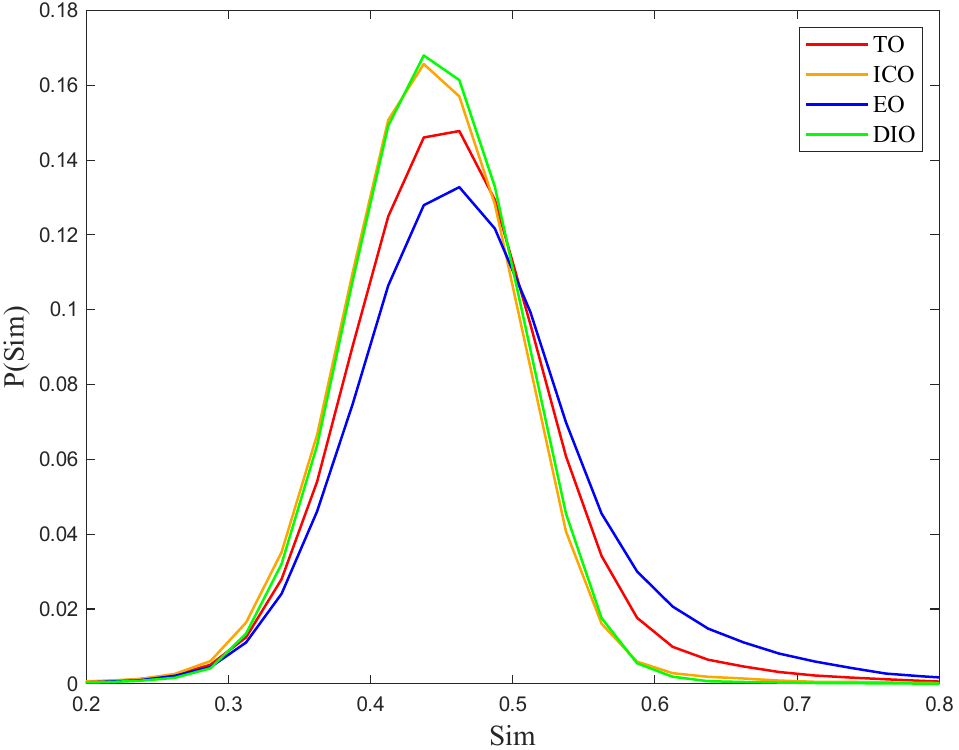}\\
  \small (b)
\end{minipage}
\caption{
    (a) Deviation distribution of daily activity times for users with different interest preferences compared to the overall average behavior of all users; (b) Probability density distribution of $E_m^{+}$.
}
\label{fig:9}
\end{figure}

In the content-oriented coupling analysis, Figure~\ref{fig:9} illustrates the behavioral differences among various interest preference groups. Figure~\ref{fig:9}(a) depicts the variations in news click frequency at the macro temporal scale across different user preference groups, relative to the overall average click frequency. The results reveal that timeliness-oriented users exhibit significantly higher click frequencies in the early morning and afternoon, while entertainment-oriented users have peak click frequencies in the afternoon and evening. Additionally, at the mesoscale, users with in-depth content, timeliness, and entertainment-oriented preferences show significantly higher click frequencies in a short period, reflecting a stronger focus on timeliness. In contrast, diverse interest-oriented users have a lower click frequency in a short time, indicating a slower information consumption pace. At the microscale, timeliness-oriented and in-depth content-oriented users display concentrated initial clicks and a fast interaction rhythm, while entertainment-oriented users have longer click intervals and a slower overall interaction pace. Users with diverse interest preferences show a lower overall deviation, with behavior patterns closer to the group average (details can be found in Supplementary Figure A8). To verify the generalizability of these conclusions, we further validated them using the MIND dataset, with results shown in Supplementary Figure A9.

In the content dimension, Figure~\ref{fig:9}(b) shows the probability density distributions of semantic similarity between clicked news items and historical news items for different preference groups. The results show that entertainment-oriented and timeliness-oriented users tend to click on content highly related to their historical interests. In contrast, the similarity distributions of In-Depth content orientation users and diverse interest-oriented users are more similar. These two preference groups are less influenced by historical interests. Additionally, we analyzed the impact of exposure diversity on groups with different preferences using the MIND dataset. We found that as exposure diversity increases, entertainment-oriented and timeliness-oriented users still maintain a preference for content highly related to their historical interests, indicating that their click behavior is strongly dependent on existing interests. In contrast, diverse interest-oriented users are more sensitive to changes in exposure content. As exposure diversity increases, the trend of their click similarity distribution shifting left becomes more significant (details can be found in Supplementary Figure A10).

In the time-oriented coupling analysis, users were classified into daytime-active and nighttime-active groups based on their 24-hour activity patterns. Statistical analyses revealed no significant differences between these groups in temporal behavior patterns or content selection mechanisms, indicating that user news consumption is predominantly influenced by individual content preferences rather than by their temporal activity profiles (see Supplementary Figure A11).

\section{Conclusion}

This study proposes a temporal-content coupling user behavior analysis framework that integrates multi-scale temporal dynamic features and content selection mechanisms. It is empirically validated using two large-scale datasets, MIND and Adressa, thereby demonstrating its validity and robustness in capturing complex user behavior patterns across temporal and content dimensions. Empirical results indicate that, at the macroscopic scale, users’ circadian rhythms within the 24-hour cycle are modeled using Fourier series, capturing the periodicity of news consumption; at the mesoscale, session intervals follow a bimodal distribution, with short intervals conforming to a power-law distribution ($\alpha \approx 1$) and long intervals synchronized with nighttime rest periods; at the microscale, both the number of actions within a session and the time intervals between actions follow exponential distributions with $\lambda \approx 0.3$ and $\lambda \approx 0.02$, respectively. In the content selection mechanism analysis, user click behavior is driven by historical interests, with the semantic similarity between clicked content and historical interests being generally higher than that of non-clicked content. Moreover, as the diversity of exposed content increases, the influence of historical interests on click behavior diminishes. Additionally, our random click experiment on Google News demonstrates that this trend is mainly attributed to user behavior rather than the recommendation system. Through the temporal-content coupling analysis, it is found that user news consumption behavior is mainly driven by content interests rather than active time periods. Differences exist in behavioral rhythm and selection mechanisms across different interest preference groups: timeliness-oriented and entertainment-oriented users exhibit a compact interaction rhythm and strong interest dependence, while diverse interest-oriented users are more sensitive to exposure changes. This study further extends the analysis method to the individual behavior level, finding that the time dynamic features and content selection mechanisms also apply at the individual scale. In addition to the temporal-content coupling analysis, this study also employs an agent-based modeling (ABM) approach to model users' behavior trajectories over a 24-hour cycle, further validating the generalizability and applicability of the findings across different user groups. By integrating three temporal scales with content selection mechanism analysis, this study overcomes the limitations of traditional temporal-content separated modeling and proposes a unified time-content coupling framework. This provides theoretical support for multi-scale user behavior understanding and empirical evidence for personalized recommendation system optimization and AI agent dynamic adaptation mechanism design. However, the current work mainly focuses on exposure content diversity, and future research could expand the analysis to more exposure features, such as content source diversity and presentation formats (e.g., text and images vs. videos), to gain a comprehensive understanding of user interaction behavior in complex environments.

\section*{Disclosure statement}

The authors declare that they have no known competing financial interests or personal relationships that could have appeared to influence the work reported in this paper.

\section*{Funding}

Funding: This work was supported by the General Project of Ministry of Education Foundation on Humanities and Social Sciences (23YJA860011); the Fundamental Research Funds for the Central Universities (1243200012); and the Guangdong Philosophy and Social Science Foundation Regular Project (GD24XXW02).

\section*{CRediT Roles}

\textbf{Jipeng Tan:} Writing – Original Draft, Writing – Review \& Editing, Visualization, Methodology, Data Curation, Formal Analysis. \textbf{Mengye Yang:} Writing – Review \& Editing, Visualization, Methodology, Formal Analysis. \textbf{Zhanghao Li:} Writing – Review \& Editing, Supervision, Methodology, Formal Analysis. \textbf{Yong Min:} Writing – Review \& Editing, Conceptualization, Validation, Supervision, Resources, Project Administration, Funding Acquisition.

\section*{Data availability statement}

The Adressa Dataset that supports the findings of this study is openly available in SmartMedia's web pages at NTNU at http://doi.org/10.1145/3106426.3109436, and the MIND Dataset that supports the findings is openly available in GitHub (via https://msnews.github.io) at http://doi.org/10.18653/v1/2020.acl-main.331.

\section{Appendices}

\appendix

\section{Random Click Behavior Experiment}

Section 3.2 of the main text has verified the semantic similarity distribution, revealing that real user click behavior exhibits a historical interest dependency characteristic (i.e., the "right-skewed phenomenon"), where the semantic similarity between clicked content and historical interests is concentrated in the higher-value range, compared to the exposed content. However, this phenomenon may arise from either the user's behavior or be influenced by the recommendation system's algorithms, such as collaborative filtering and content-based recommendation mechanisms. To clarify the attribution, we employ a social robot field experiment to simulate unbiased random click behavior, constructing a controlled scenario to compare with real user behavior and systematically validate the true source of the right-skewed phenomenon.

We selected 36 newly registered Google News accounts with no browsing history, focusing on the platform’s "For You" personalized recommendation page. Through a standardized random exposure and click process, we created a contrast scenario to that in Section 3.2. Specifically, during each active session, the robot account loaded 50 recommended news articles as the exposure set, randomly clicking 10 articles with equal probability and mimicking human user stay behavior. Metadata such as news titles, sources, and timestamps were recorded concurrently. The experiment ran from September 1 to 30, 2023, with activity at four non-peak times: 00:00, 06:00, 12:00, and 18:00, which strengthened the control over experimental behavior by fixing activity times. A total of 181,694 data records were collected, with an average of 5,047 records per account, the highest being 5,144, the lowest 4,709, and a standard deviation of 140, indicating a balanced data distribution.

To build a comparable historical interest sequence to real user behavior, we followed the MIND dataset processing methodology, using a cumulative summation approach to generate user history: the historical sequence for the nth experiment consists of the click content from the previous $n-1$ experiments (for example, the history for the 2nd experiment is the click sequence from the 1st experiment, and the history for the 3rd experiment is the union of the first two click sequences). Ultimately, the last 50 exposure units for each account were selected for analysis.

Based on the semantic similarity analysis framework in Section 3.2 of the main text, we analyze the data from the random click experiment. The results of the analysis are shown in Supplementary Fig.~\ref{fig:A12}. The entropy value ($En$) is used to distinguish the diversity of exposed content and measure the similarity between news and historical sequences. The distribution of data points for blue (unclicked exposed set, $E^-$) and red (randomly clicked exposure, $E^+$) shows that, compared to unclicked exposures, randomly clicked content does not exhibit a higher similarity. Furthermore, as exposure diversity increases, the median similarity within each entropy interval remains stable, indicating that exposure diversity does not significantly reduce the historical dependency of random clicks. This result directly supports the conclusion that the right-skewed phenomenon in real user clicks fundamentally stems from human active interest dependency, rather than interference from the mechanisms of recommendation system algorithms.

\section*{Supplementary Figures}
\begin{figure}[htbp]
    \centering
    \includegraphics[width=0.6\textwidth]{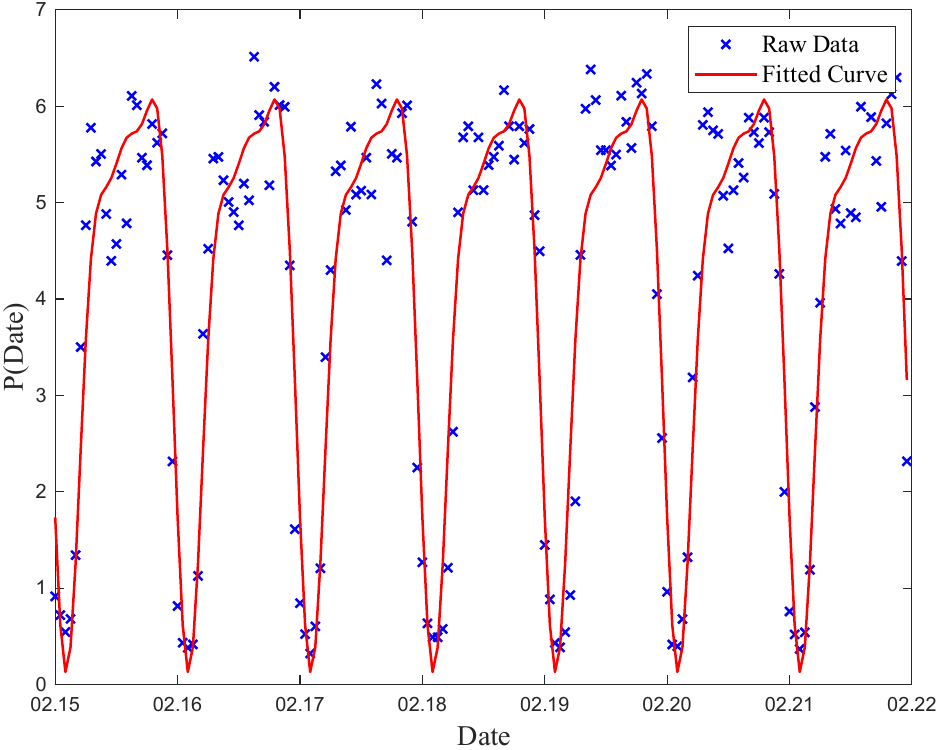}
    \caption{Scatter plots and Fourier fitting curves randomly selected from data collected over a period of 7 days.}
    \label{fig:A1}
\end{figure}
\begin{figure}[htbp]
    \centering
    
    \begin{subfigure}[b]{0.49\textwidth}
      \centering
      \includegraphics[width=\textwidth]{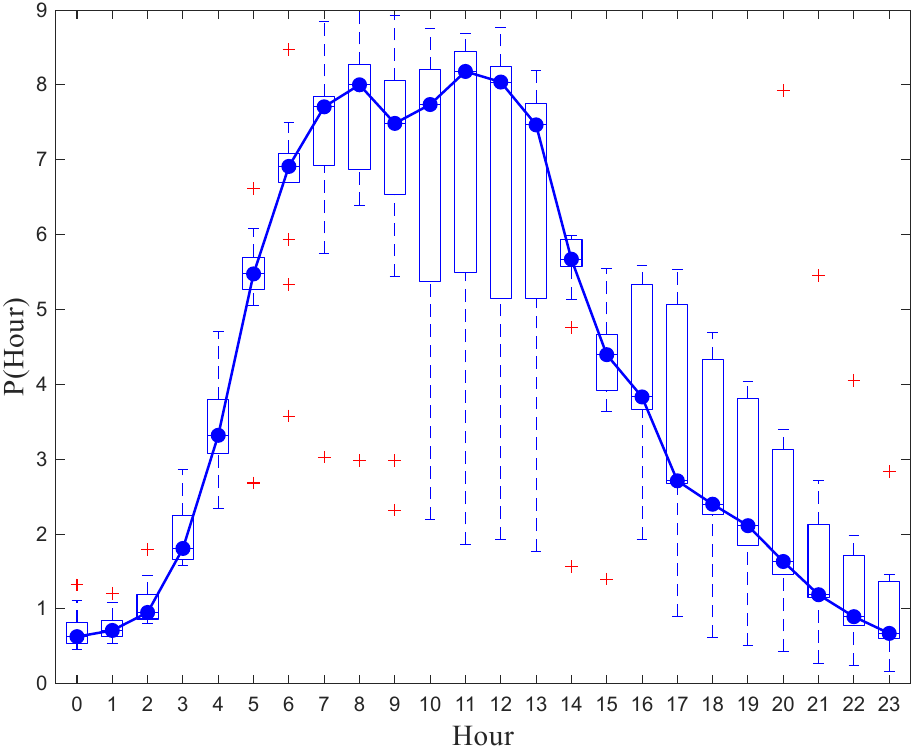}
      \caption{}
      \label{fig:A2a}
    \end{subfigure}
    \hfill
    \begin{subfigure}[b]{0.49\textwidth}
      \centering
      \includegraphics[width=\textwidth]{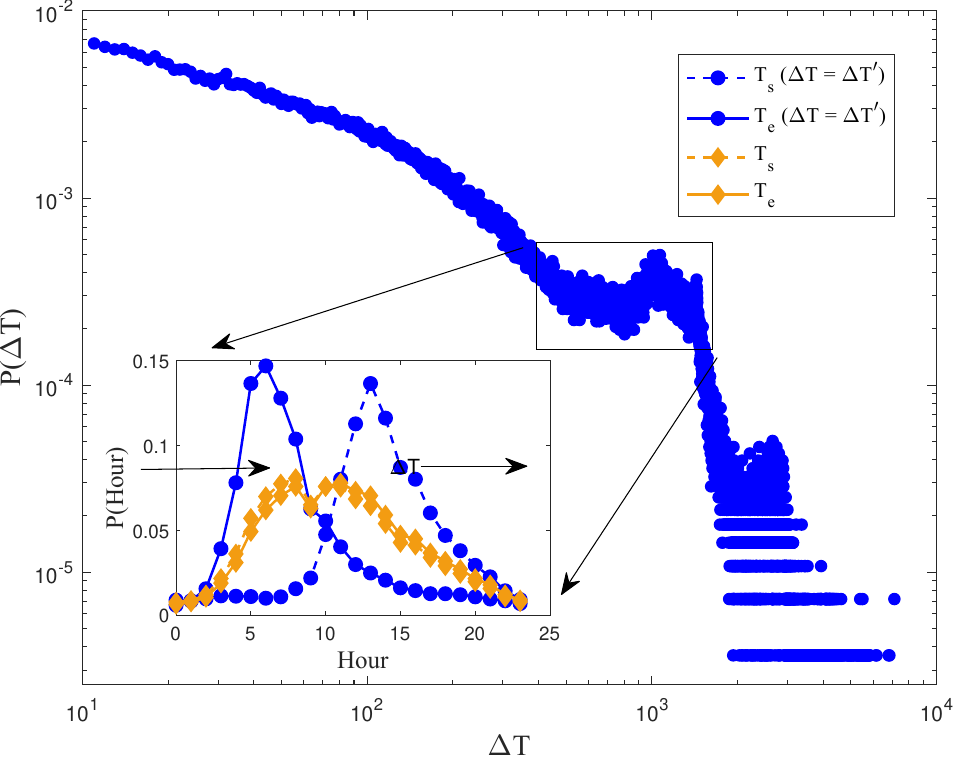}
      \caption{}
      \label{fig:A2b}
    \end{subfigure}
    \label{fig:A2}
\end{figure}

\clearpage

\begin{figure}[htbp]
    \ContinuedFloat 
    \centering
    \begin{subfigure}[b]{0.49\textwidth}
      \centering
      \includegraphics[width=\textwidth]{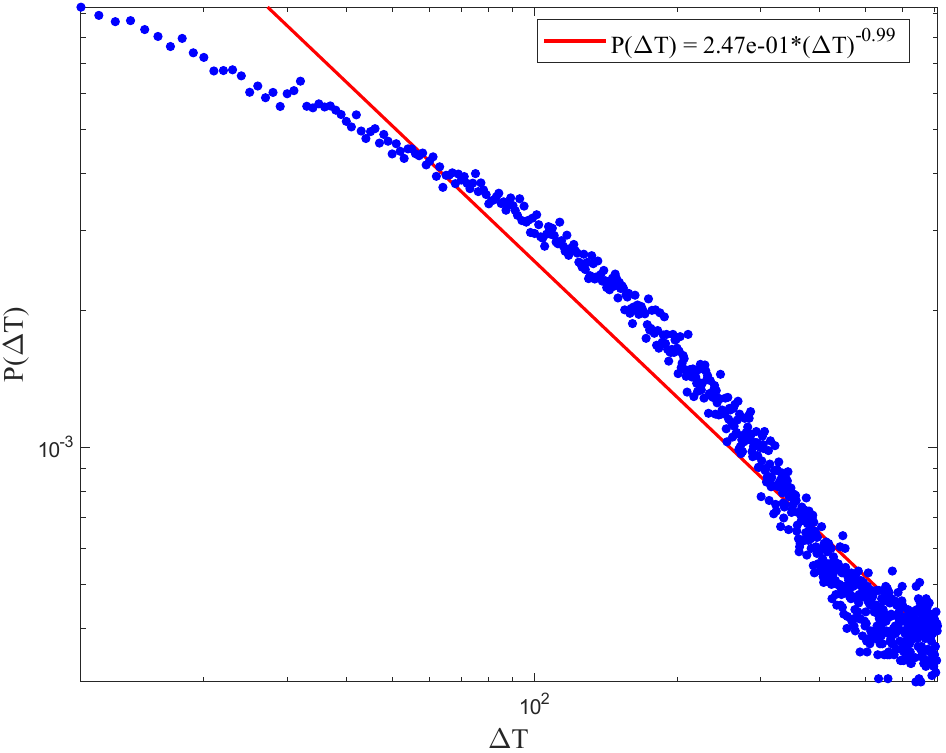}
      \caption{}
      \label{fig:A2c}
    \end{subfigure}
    \hfill
    \begin{subfigure}[b]{0.49\textwidth}
      \centering
      \includegraphics[width=\textwidth]{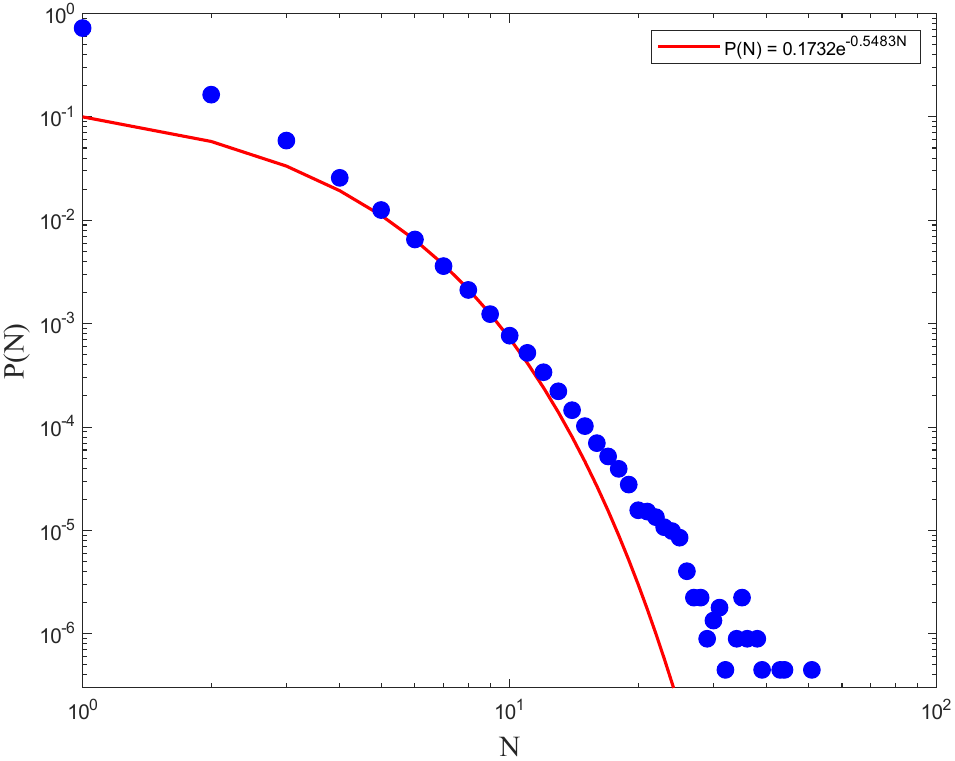}
      \caption{}
      \label{fig:A2d}
    \end{subfigure}
    \caption[]{Analysis of the MIND dataset at three time scales: (a) macroscopic level, (b) mesoscopic level, (c) power-law fit at mesoscopic level for $\Delta T < \Delta T'$, (d) exponential fit at microscopic intralayer level.}
\end{figure}


\begin{figure}[htbp]
    \centering
    \begin{subfigure}[b]{\textwidth}
        \centering
        \includegraphics[width=0.65\textwidth]{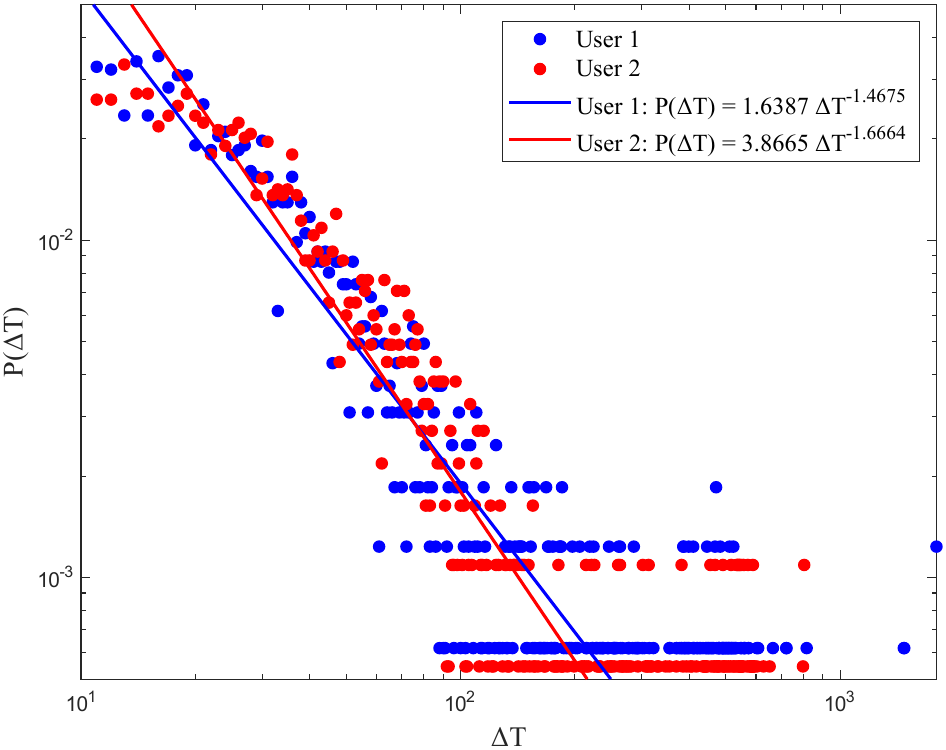}
        \caption{}
        \label{fig:A3a}
    \end{subfigure}

\end{figure}
\clearpage
\begin{figure}[htbp]    
    \ContinuedFloat 
    \begin{subfigure}[b]{\textwidth}
        \centering
        \includegraphics[width=0.65\textwidth]{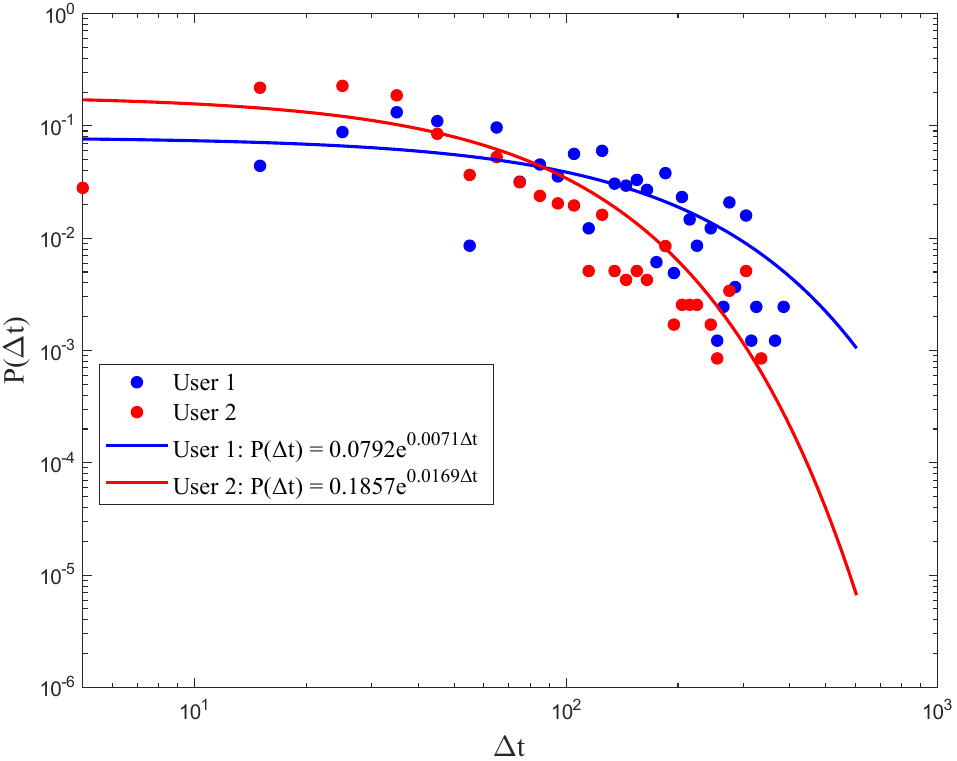}
        \caption{}
        \label{fig:A3b}
    \end{subfigure}
    
    \vspace{2em}  

    \begin{subfigure}[b]{\textwidth}
        \centering
        \includegraphics[width=0.65\textwidth]{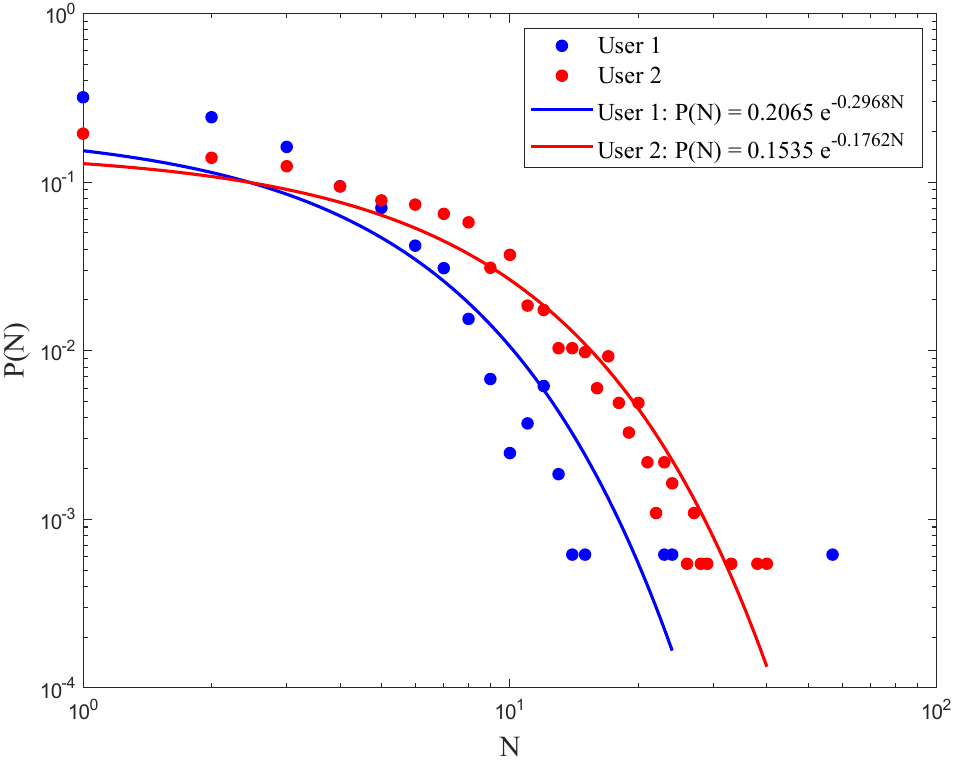}
        \caption{}
        \label{fig:A3c}
    \end{subfigure}

    \caption{Random analysis of two users from the Adressa dataset at mesoscopic and microscopic time scales: (a) empirical fit of mesoscopic session intervals, (b) empirical fit of microscopic clicks, (c) empirical fit of microscopic operation time intervals.}
    \label{fig:A3}
\end{figure}


\begin{figure}[htbp]
    \centering
    \begin{subfigure}[b]{0.49\textwidth}
        \centering
        \includegraphics[width=\textwidth]{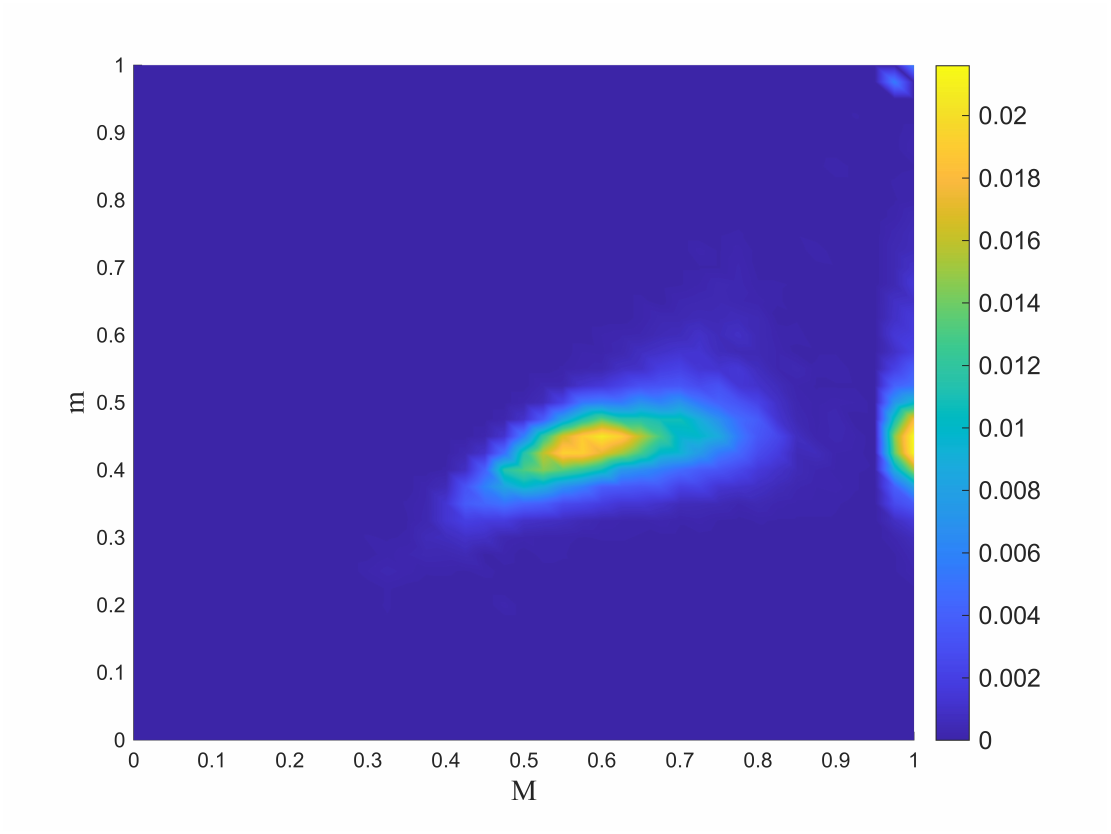}
        \caption{}
        \label{fig:A4a}
    \end{subfigure}
    \hfill
    \begin{subfigure}[b]{0.49\textwidth}
        \centering
        \includegraphics[width=\textwidth]{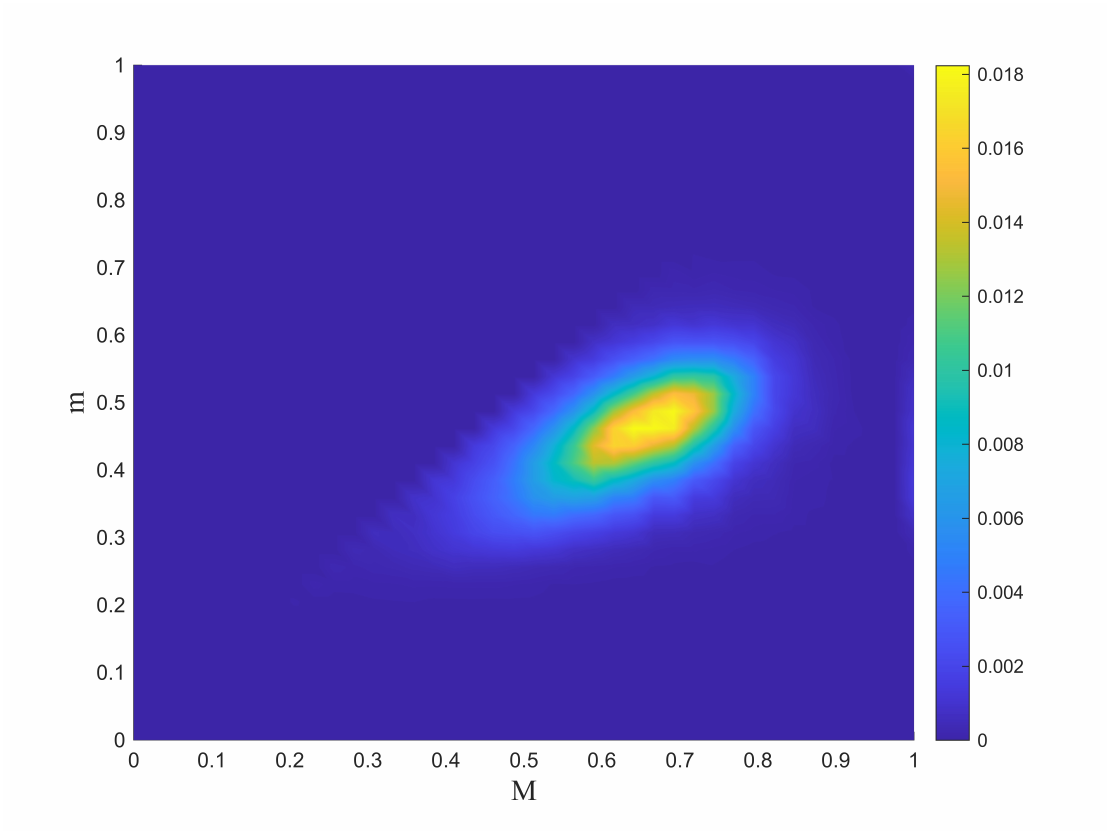}
        \caption{}
        \label{fig:A4b}
    \end{subfigure}

    \caption{Joint probability density distributions of $E_M^+$ and $E_m^+$ for different datasets: (a) Adressa, (b) MIND. In both datasets, the similarity between clicked news and historical records is concentrated in higher ranges.}
    \label{fig:A4}
\end{figure}


\begin{figure}[htbp]
    \centering
    \includegraphics[width=0.65\textwidth]{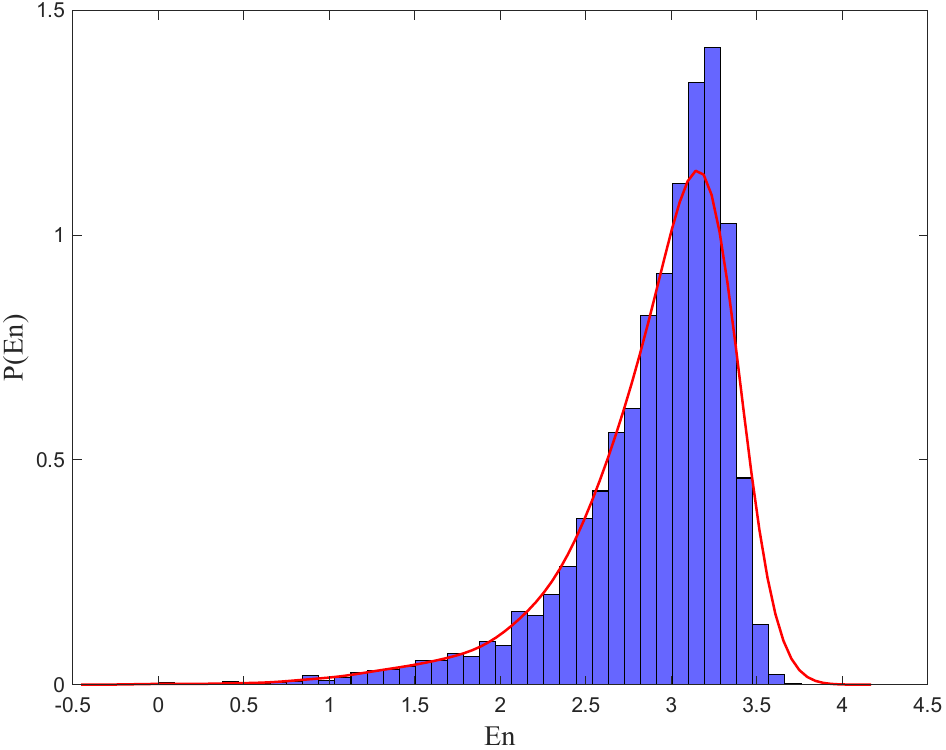}
    \caption{The function of $P(E_n)$ with respect to $E_n$ shows that most values of $a$ fall between 1.2 and 3.6.}
    \label{fig:A5}
\end{figure}


\begin{figure}[htbp]
    \centering
    \begin{subfigure}[b]{\textwidth}
        \centering
        \includegraphics[width=0.65\textwidth]{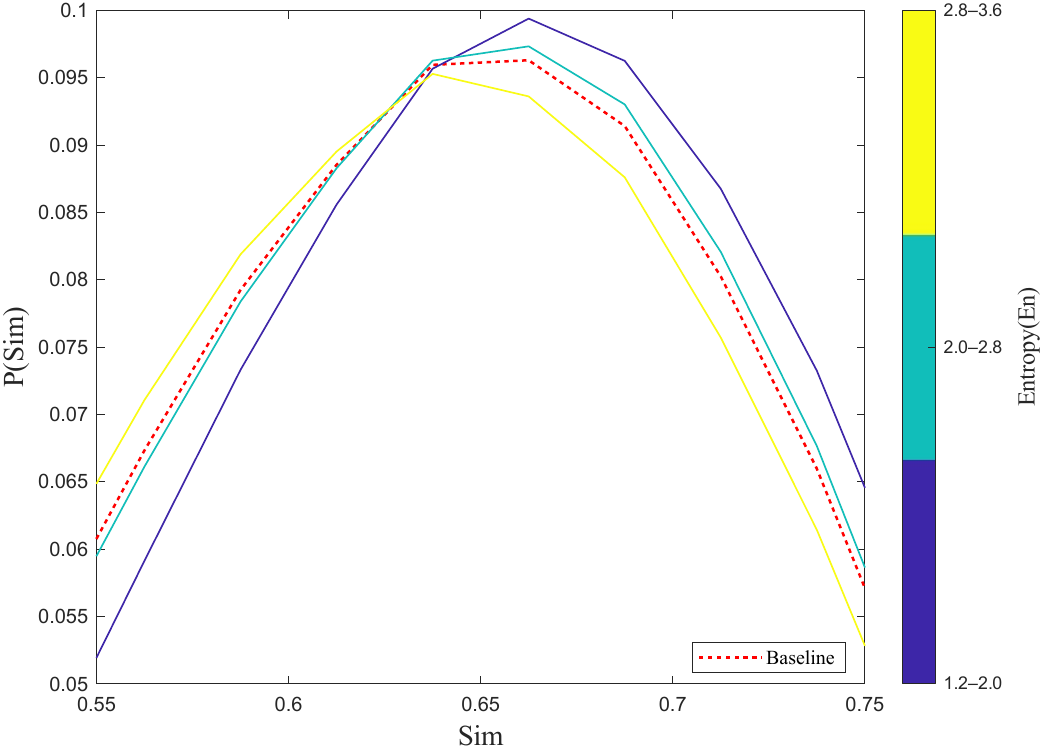}
        \caption{}
        \label{fig:A6a}
    \end{subfigure}

    \vspace{2em}

    \begin{subfigure}[b]{\textwidth}
        \centering
        \includegraphics[width=0.65\textwidth]{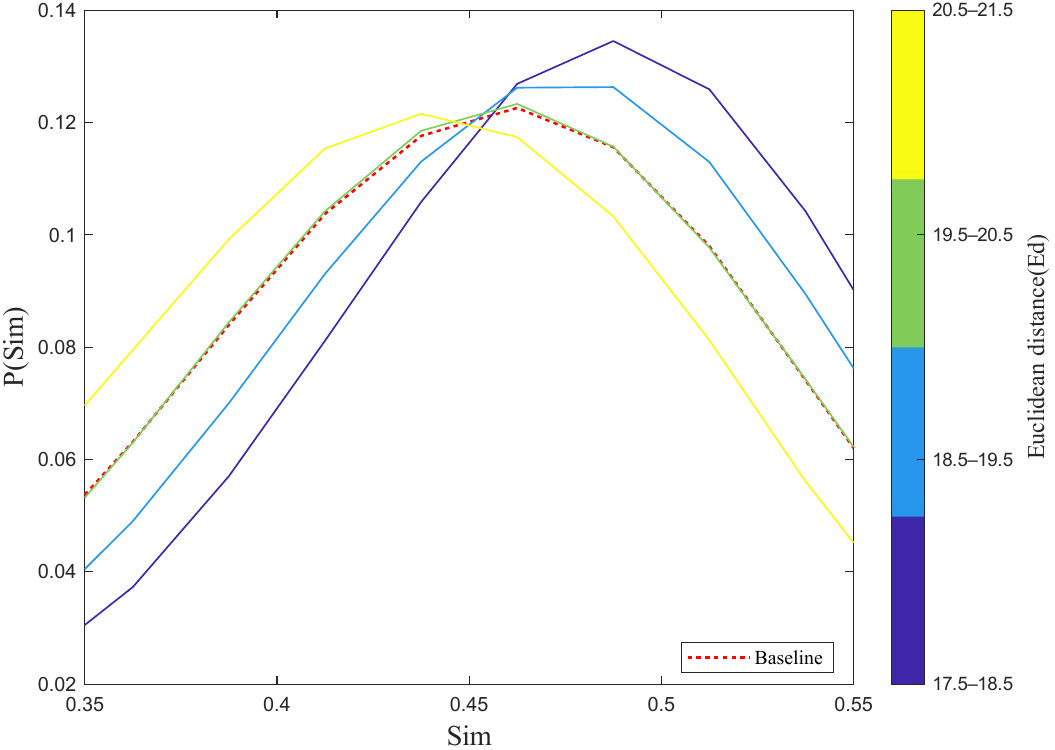}
        \caption{}
        \label{fig:A6b}
    \end{subfigure}

\end{figure}
\clearpage
\begin{figure}[htbp] 
    \ContinuedFloat 
    \begin{subfigure}[b]{\textwidth}
        \centering
        \includegraphics[width=0.65\textwidth]{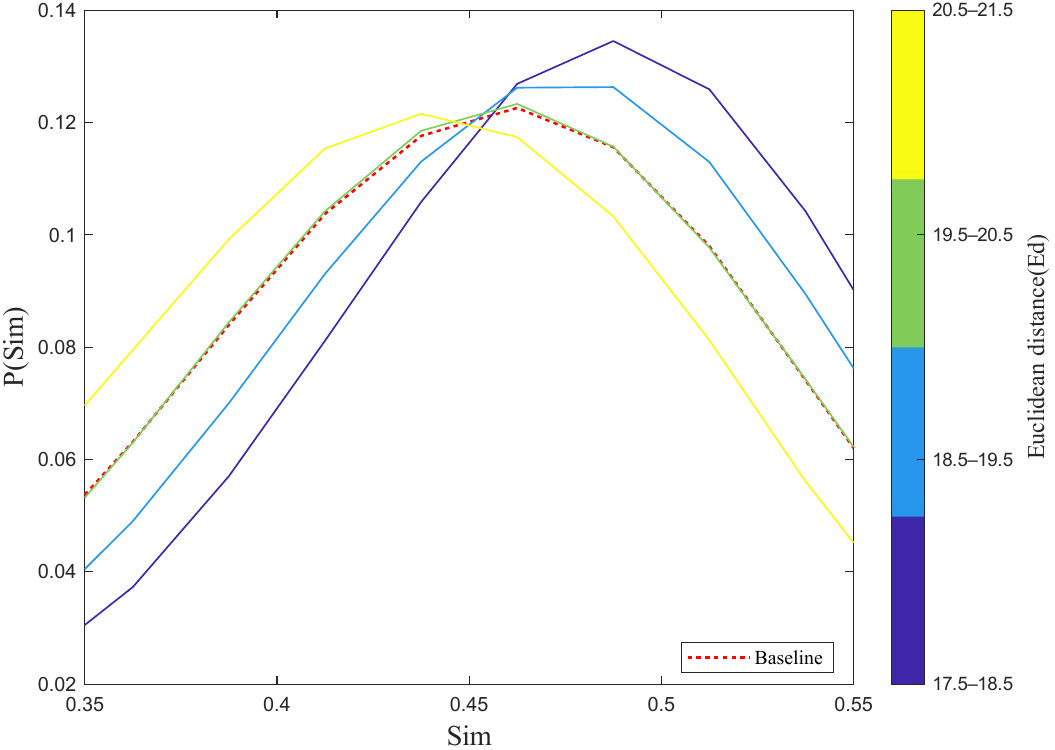}
        \caption{}
        \label{fig:A6c}
    \end{subfigure}

    \caption{(a) Distribution differences of $E_M^+$ and Euclidean distance evaluation under varying exposure list diversities, illustrating the impact of exposure sequence diversity. (b) Distribution differences of $E_m^+$. (c) Distribution differences of $E_M^+$, both showing a leftward shift with increasing exposure list diversity. The similarity between clicks and history decreases, indicating a gradual reduction in the influence of historical interests on users' news clicking behavior.}
    \label{fig:A6}
\end{figure}


\begin{figure}[htbp]
    \centering
    \begin{subfigure}[b]{0.45\textwidth}
        \centering
        \includegraphics[width=\textwidth]{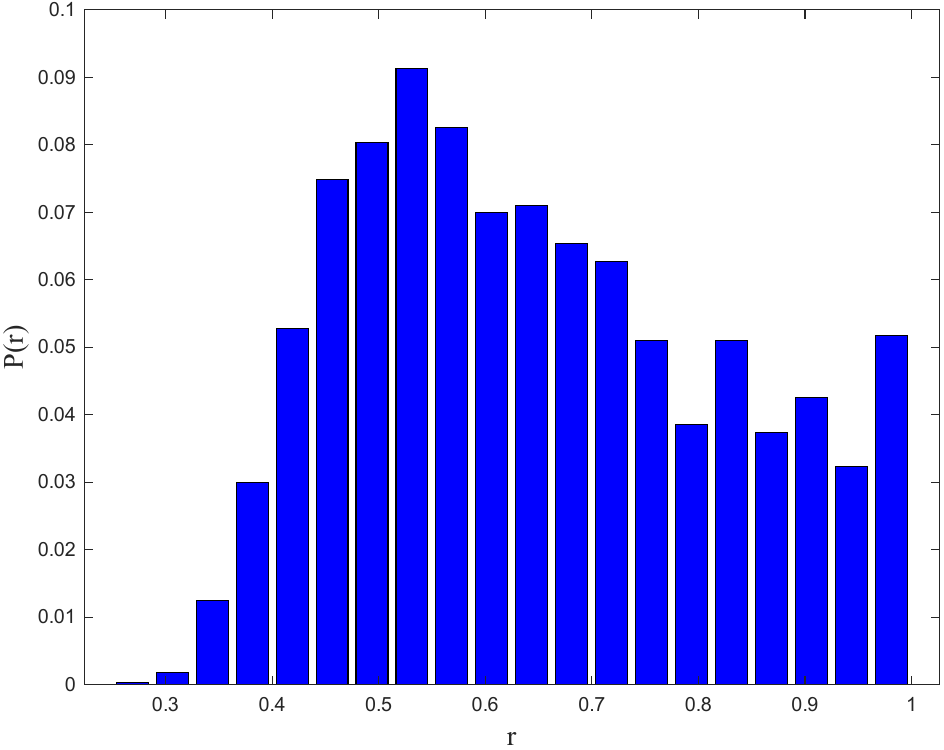}
        \caption{}
        \label{fig:A7a}
    \end{subfigure}
    \hfill
    \begin{subfigure}[b]{0.45\textwidth}
        \centering
        \includegraphics[width=\textwidth]{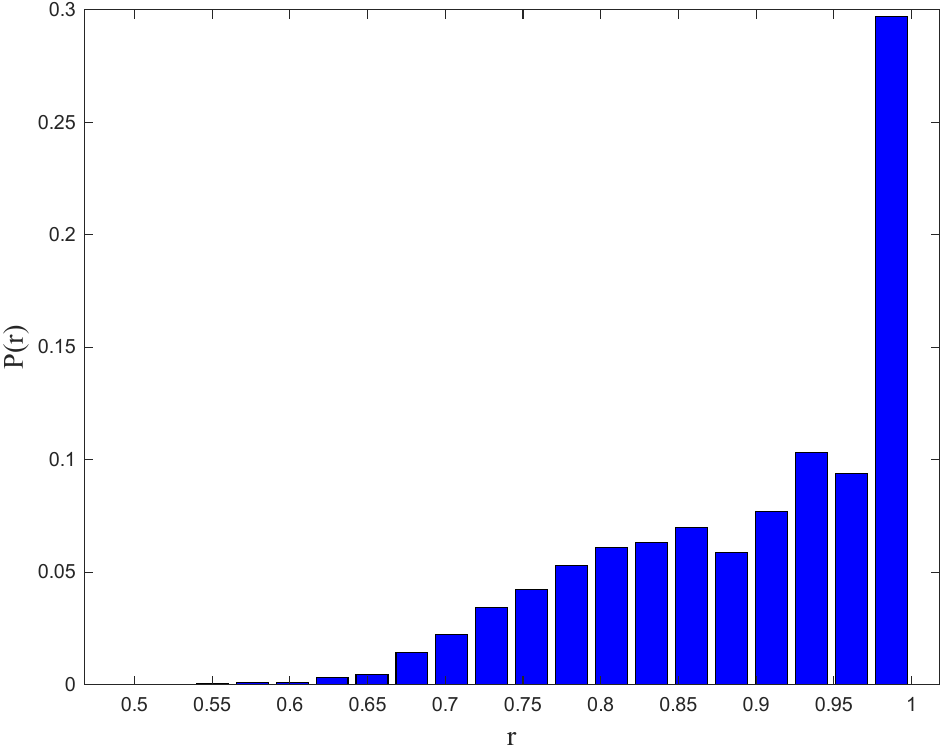}
        \caption{}
        \label{fig:A7b}
    \end{subfigure}

\end{figure}
\clearpage
\begin{figure}[htbp] 

    \ContinuedFloat 
    \begin{subfigure}[b]{0.45\textwidth}
        \centering
        \includegraphics[width=\textwidth]{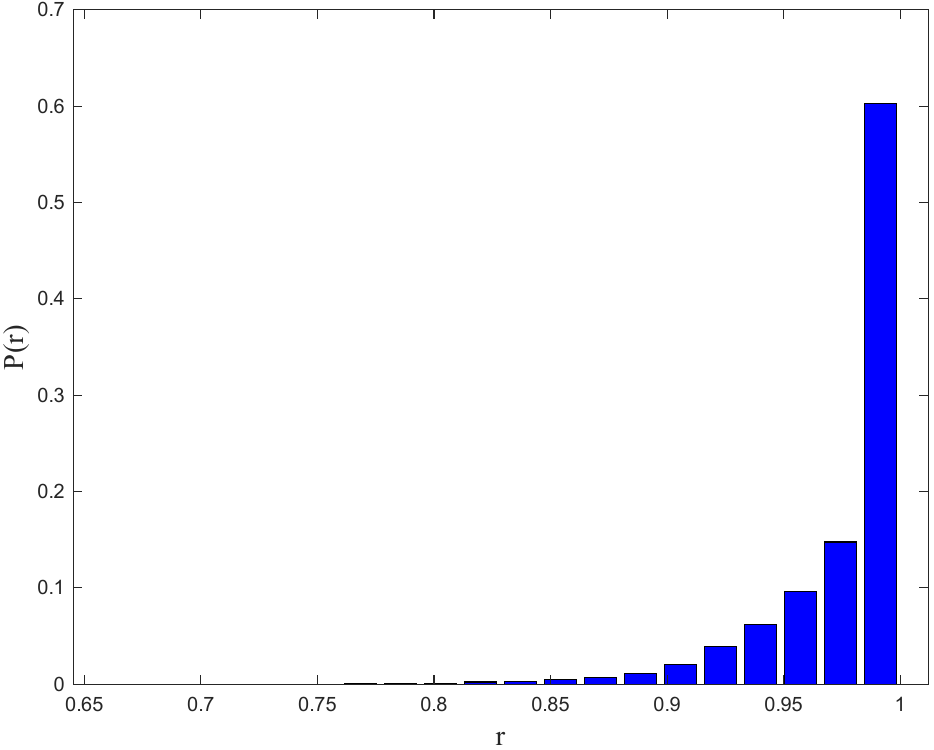}
        \caption{}
        \label{fig:A7c}
    \end{subfigure}
    \hfill
    \begin{subfigure}[b]{0.45\textwidth}
        \centering
        \includegraphics[width=\textwidth]{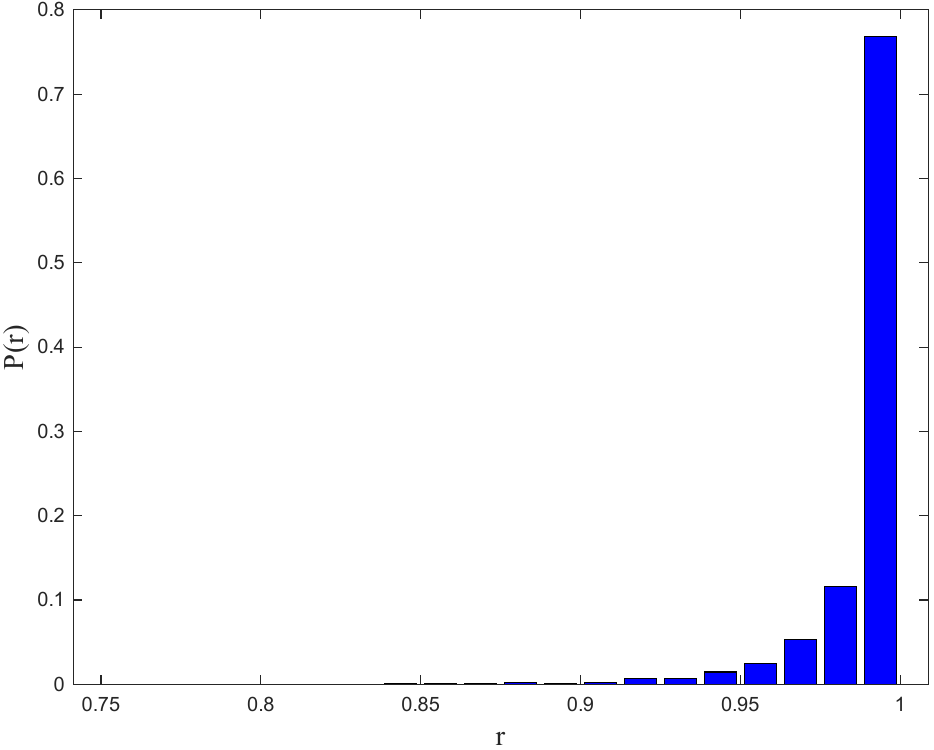}
        \caption{}
        \label{fig:A7d}
    \end{subfigure}

    \caption{To effectively extract users' interest preference features, we computed the frequency distribution of clicked categories and the coverage ratio of dominant categories. As illustrated in (a), the majority of users concentrate their clicks within 2 to 4 categories, with the top three categories covering over 90\% of total clicks. Hence, selecting the top three most frequently clicked categories per user captures the core preferences while reducing model complexity, thereby ensuring good representativeness and generalization.}
    \label{fig:A7}
\end{figure}


\begin{figure}[htbp]
    \centering
    \begin{subfigure}[b]{\textwidth}
        \centering
        \includegraphics[width=0.60\textwidth]{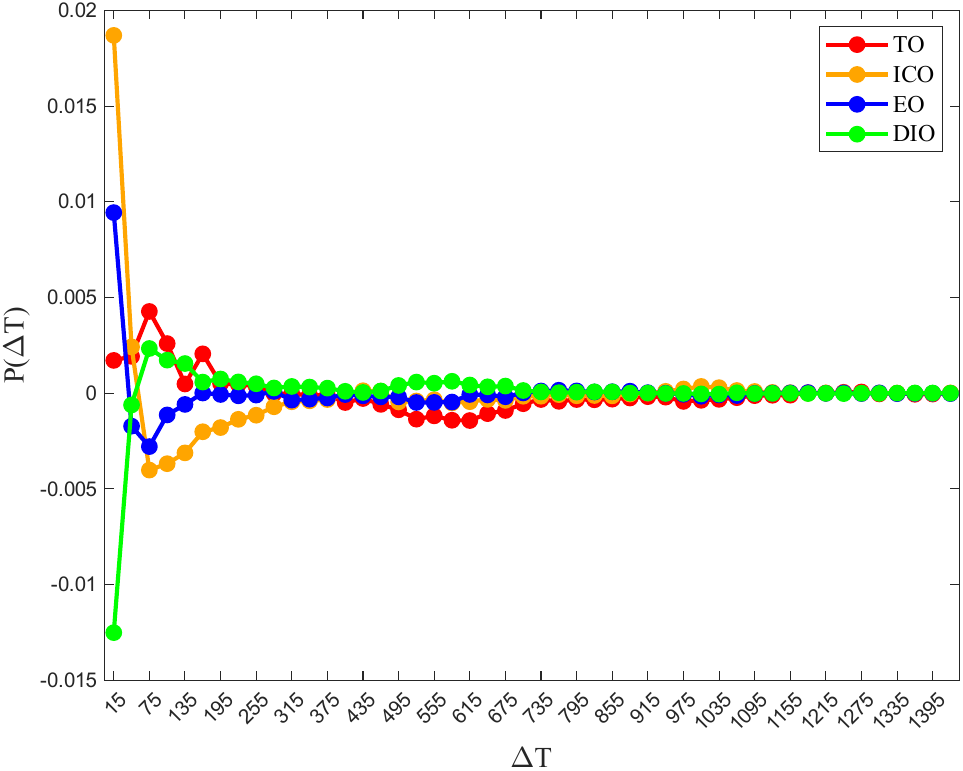}
        \caption{}
        \label{fig:A8a}
    \end{subfigure}

\end{figure}
\clearpage
\begin{figure}[htbp] 

    \ContinuedFloat 
    \begin{subfigure}[b]{\textwidth}
        \centering
        \includegraphics[width=0.65\textwidth]{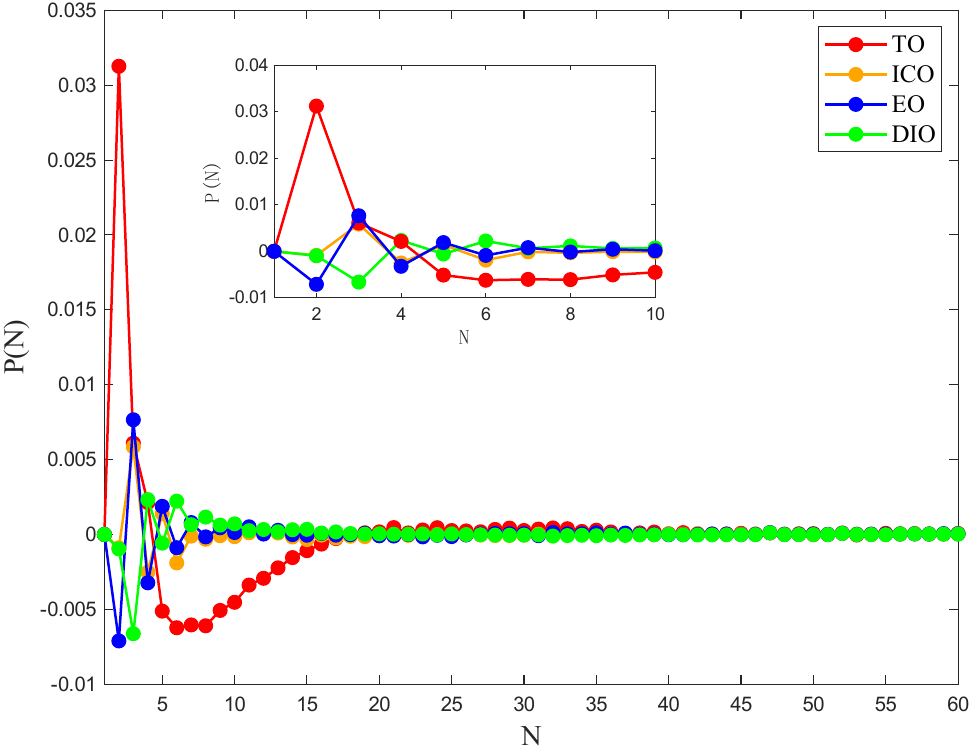}
        \caption{}
        \label{fig:A8b}
    \end{subfigure}

    \vspace{1em}

    \begin{subfigure}[b]{\textwidth}
        \centering
        \includegraphics[width=0.65\textwidth]{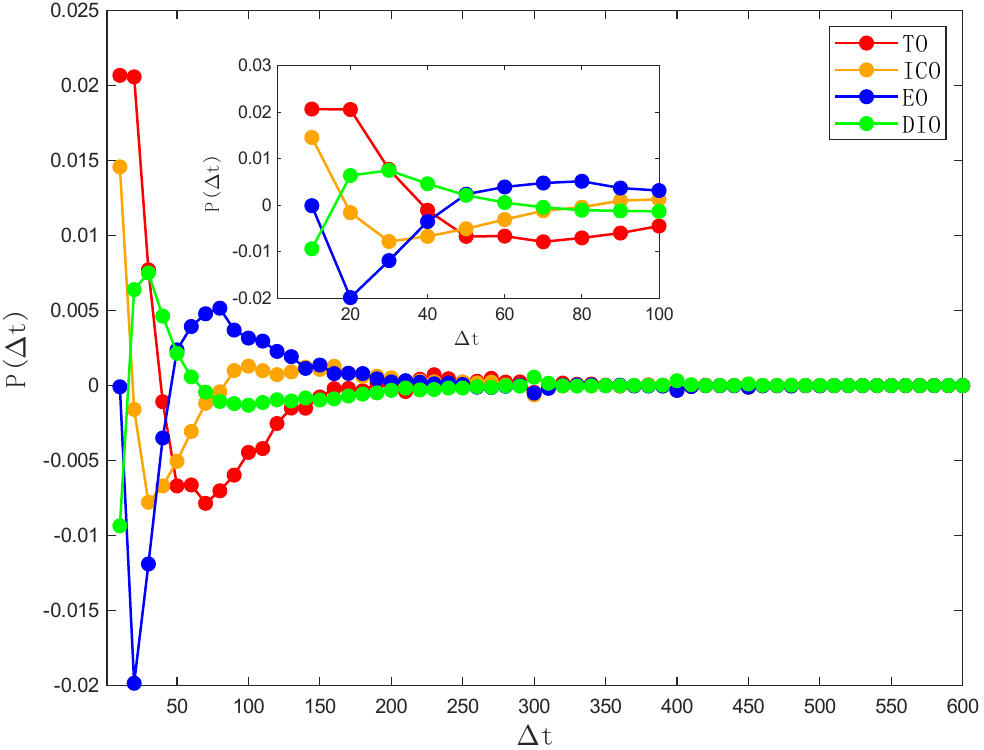}
        \caption{}
        \label{fig:A8c}
    \end{subfigure}

    \caption{Based on the Adressa dataset, (a) deviation distribution of users with different interest preferences compared to the overall user average behavior at the meso-level. (b) Deviation distribution of the number of clicks within single sessions at the micro-level. (c) Deviation distribution of time intervals between consecutive actions within single sessions at the micro-level.}
    \label{fig:A8}
\end{figure}


\begin{figure}[htbp]
    \centering
    \begin{subfigure}[b]{\textwidth}
        \centering
        \includegraphics[width=0.65\textwidth]{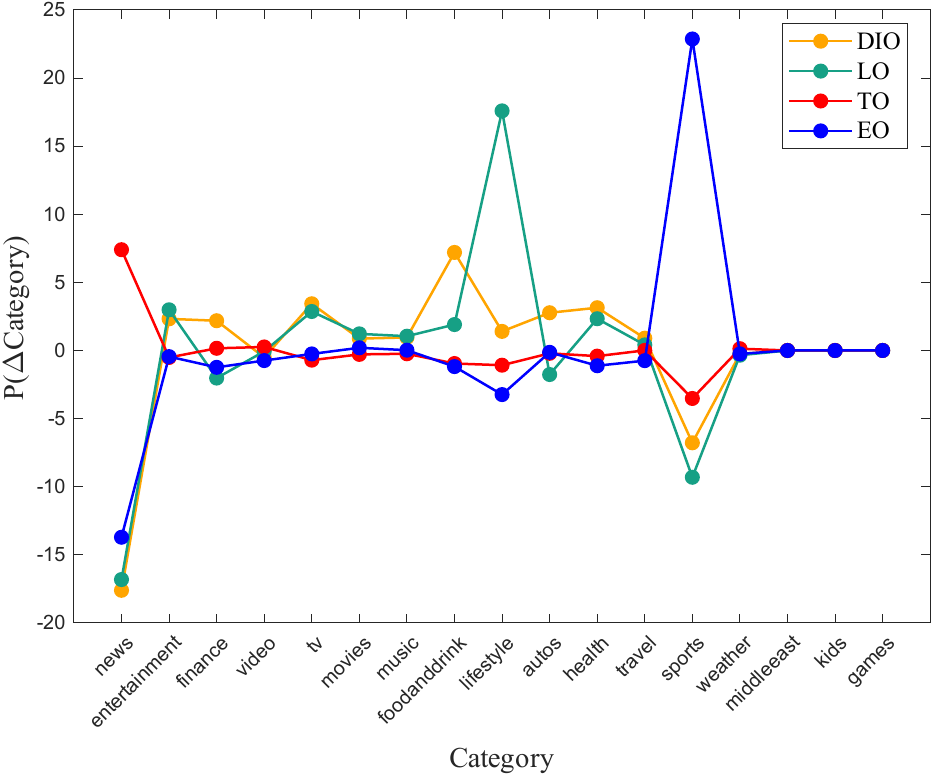}
        \caption{}
        \label{fig:A9a}
    \end{subfigure}

    \vspace{2em}

    \begin{subfigure}[b]{\textwidth}
        \centering
        \includegraphics[width=0.65\textwidth]{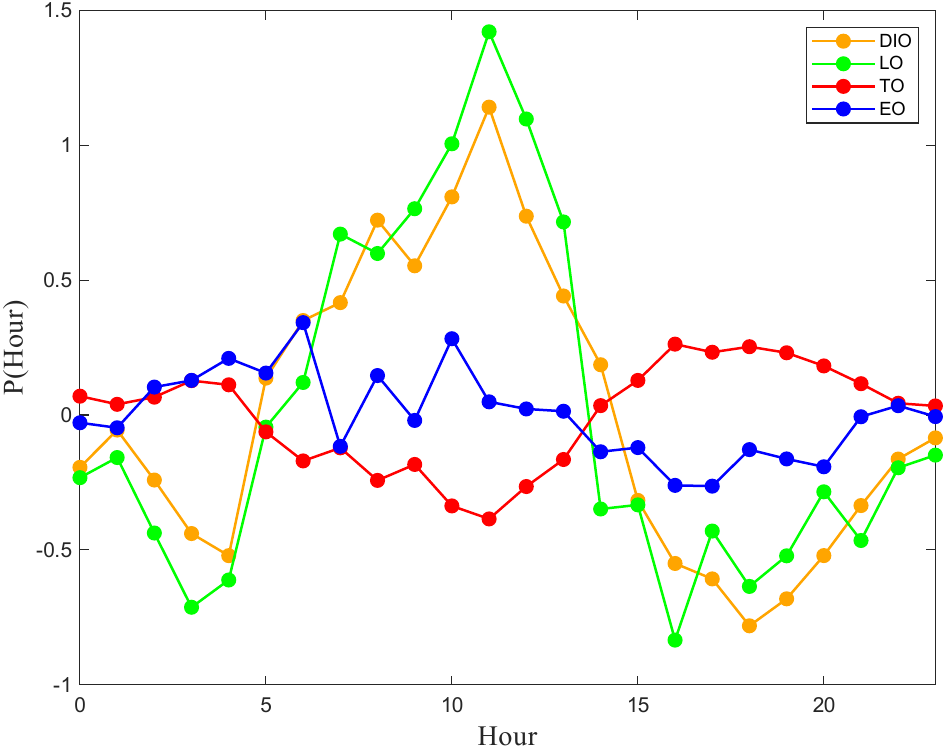}
        \caption{}
        \label{fig:A9b}
    \end{subfigure}

\end{figure}
\clearpage
\begin{figure}[htbp] 
    \ContinuedFloat 
    \begin{subfigure}[b]{\textwidth}
        \centering
        \includegraphics[width=0.65\textwidth]{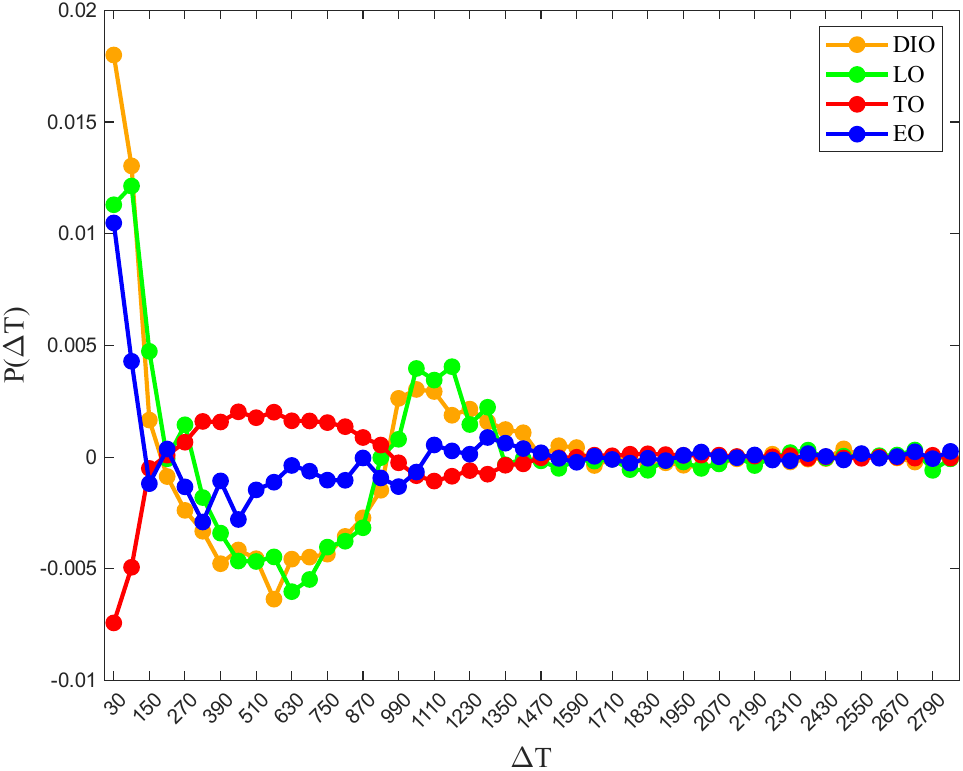}
        \caption{}
        \label{fig:A9c}
    \end{subfigure}

    \caption{Based on the MIND dataset, (a) illustrates the deviation of news-clicking frequencies among different interest preference groups relative to the overall average. Panels (b) and (c) depict the deviations at the macro and meso levels, respectively. Time-preference users exhibit relatively stable clicking rhythms; life-preference users are primarily active during daytime hours; entertainment-preference users demonstrate concentrated behavior within a narrow category range; while multi-interest users show more dispersed active periods and higher click frequencies. A comparative analysis with the Adressa dataset reveals that the MIND dataset exhibits similar macro-level temporal rhythmic patterns. Both TO and DIO users demonstrate consistent multi-phase behavioral trends across the two datasets, generally following a “decline–rise–decline (with possible rebound)” trajectory, despite some temporal shifts in peak activity periods. In contrast, EO users display markedly divergent behaviors across platforms, indicating greater sensitivity to platform-specific contexts and stronger cross-platform heterogeneity.}
    \label{fig:A9}
\end{figure}


\begin{figure}[htbp]
    \centering
    \begin{subfigure}[b]{0.45\textwidth}
        \centering
        \includegraphics[width=\textwidth]{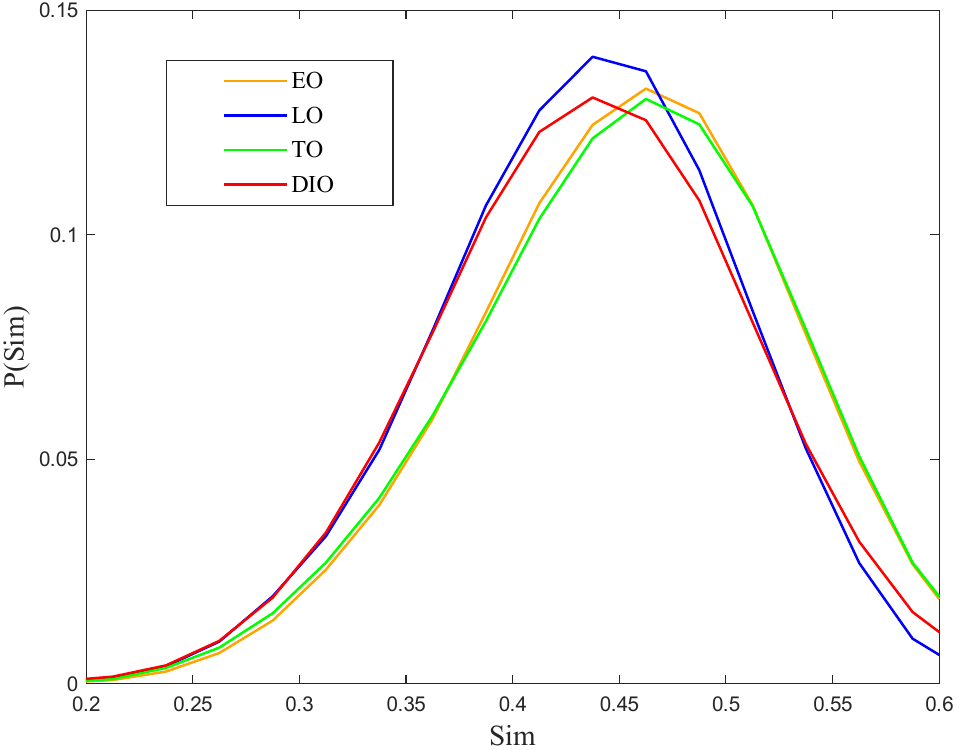}
        \caption{}
        \label{fig:A10a}
    \end{subfigure}
    \hfill
    \begin{subfigure}[b]{0.45\textwidth}
        \centering
        \includegraphics[width=\textwidth]{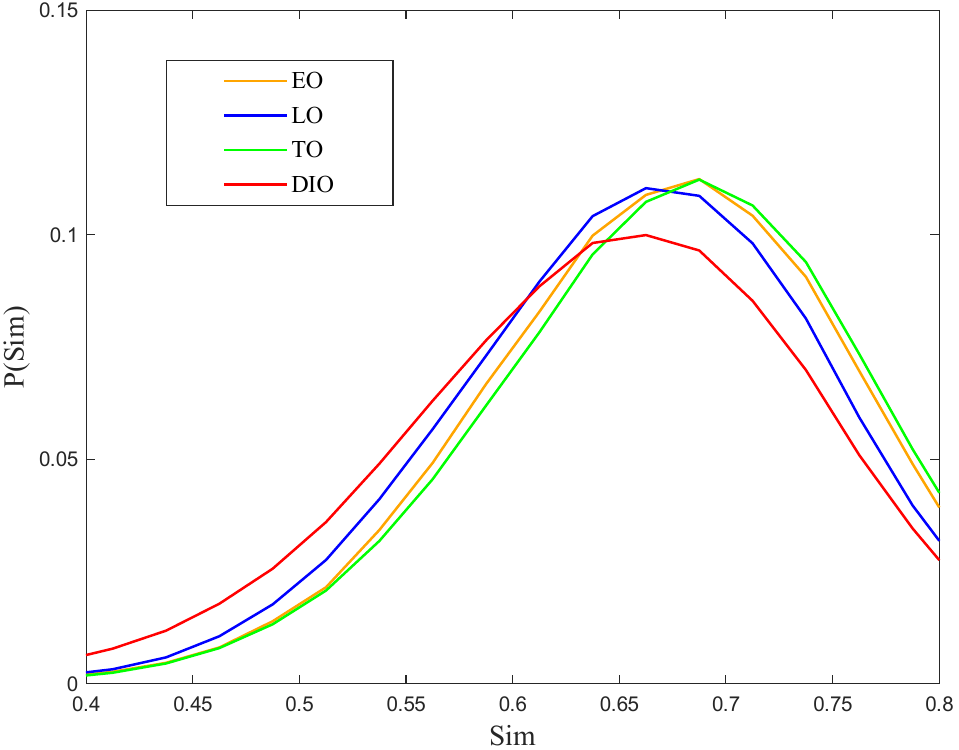}
        \caption{}
        \label{fig:A10b}
    \end{subfigure}

    \vspace{1em}

    \begin{subfigure}[b]{0.45\textwidth}
        \centering
        \includegraphics[width=\textwidth]{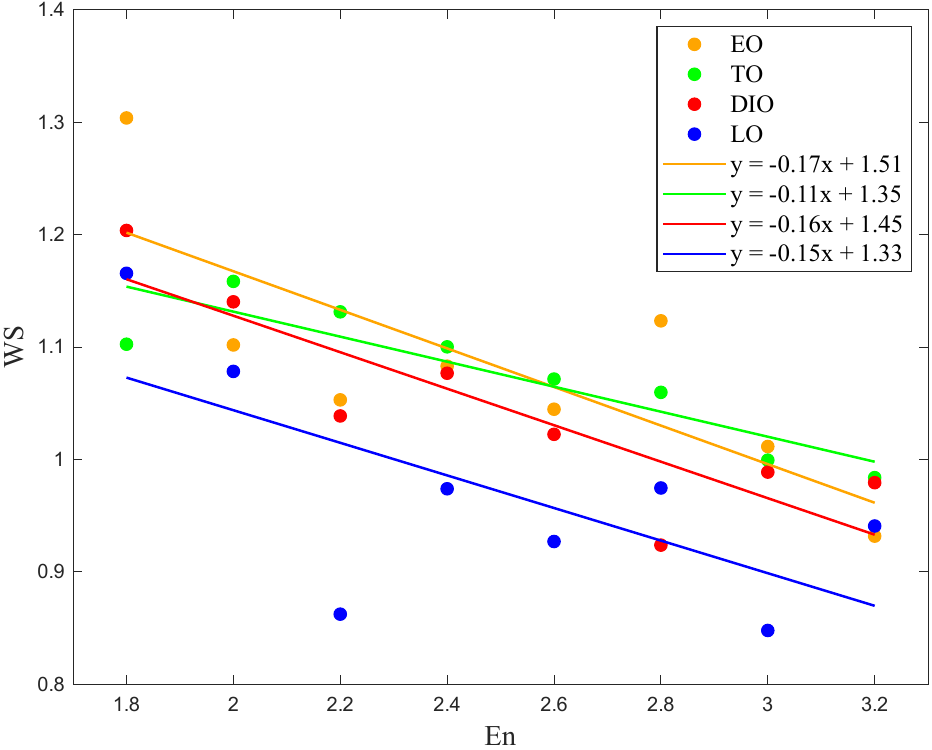}
        \caption{}
        \label{fig:A10c}
    \end{subfigure}
    \hfill
    \begin{subfigure}[b]{0.45\textwidth}
        \centering
        \includegraphics[width=\textwidth]{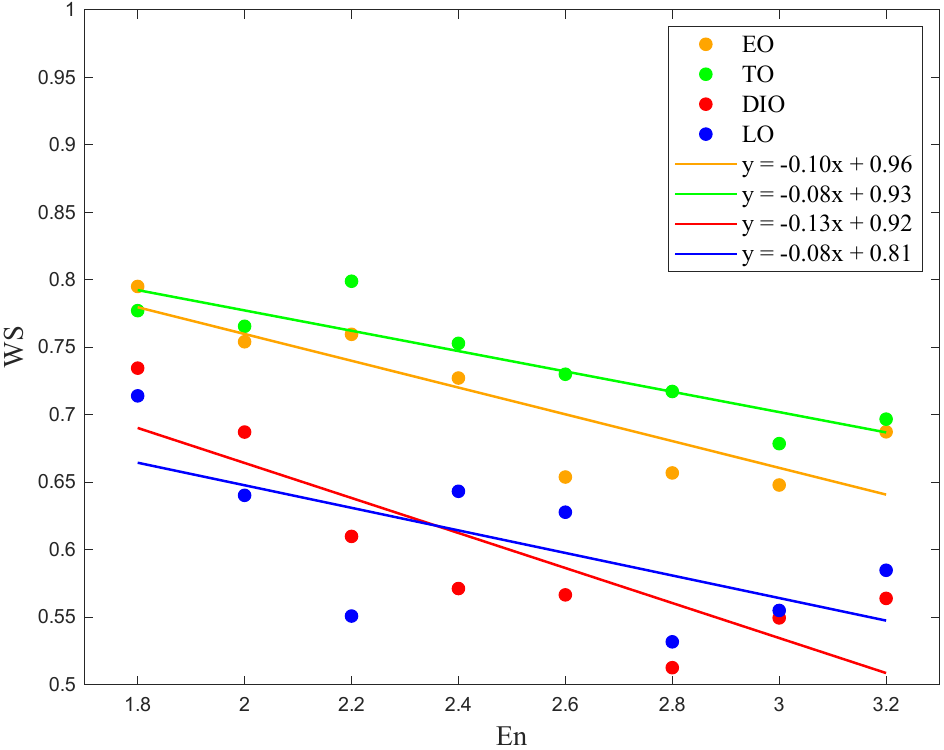}
        \caption{}
        \label{fig:A10d}
    \end{subfigure}

    \caption{(a) Probability density distribution of $E_m^+$ for different preference groups. (b) Probability density distribution of $E_M^+$. (c) Functional relationship of $W(E_m^+,E_m^-)$ with respect to $E_n$. (d) Functional relationship of $W(E_M^+,E_M^-)$ with respect to $E_n$. Compared to the Adressa dataset, the MIND dataset shows higher content similarity between clicked items and users' historical interactions for both EO and TO users, while DIO users exhibit relatively weaker similarity.}
    \label{fig:A10}
\end{figure}


\begin{figure}[htbp]
    \centering
    \begin{subfigure}[b]{0.45\textwidth}
        \centering
        \includegraphics[width=\textwidth]{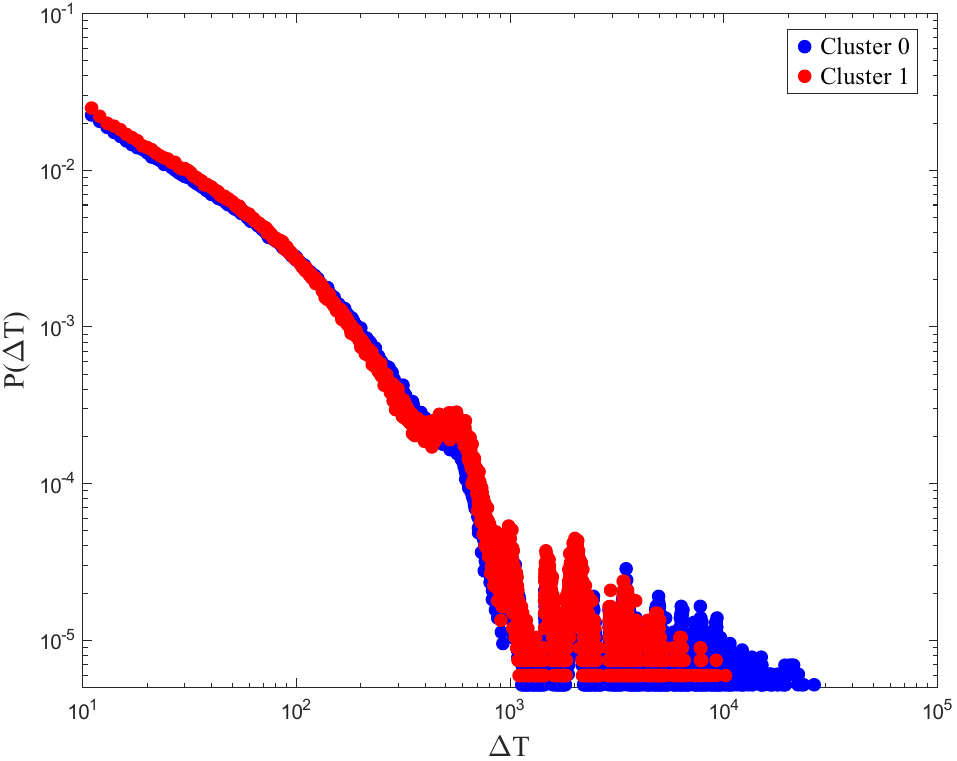}
        \caption{}
        \label{fig:A11a}
    \end{subfigure}
    \hfill
    \begin{subfigure}[b]{0.45\textwidth}
        \centering
        \includegraphics[width=\textwidth]{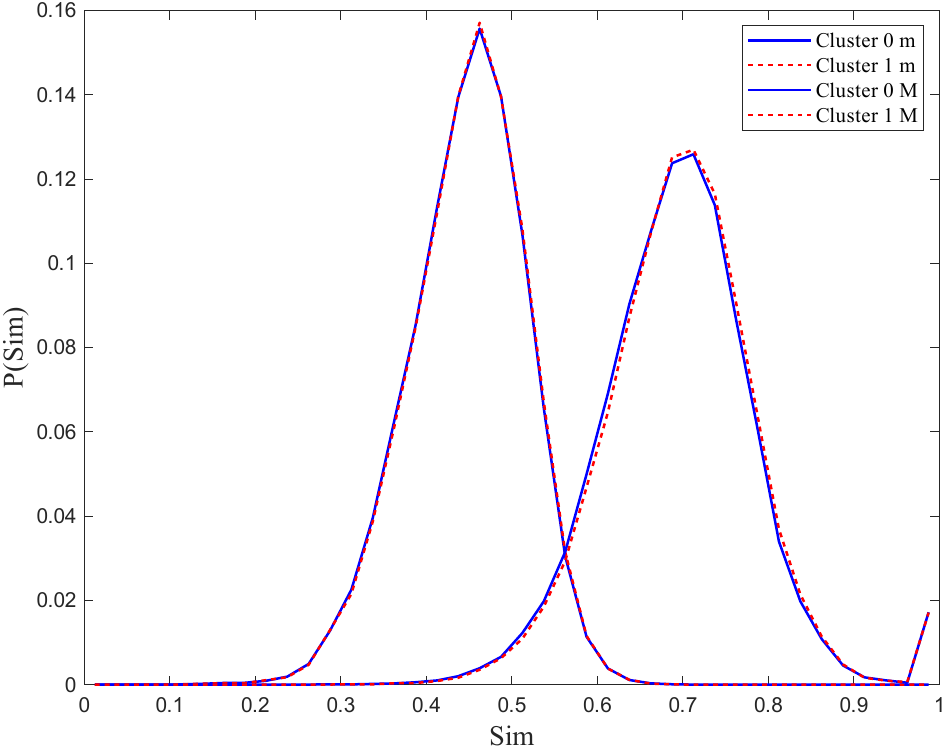}
        \caption{}
        \label{fig:A11b}
    \end{subfigure}

    \caption{Coupling analysis of time orientation: Differences among users with varying active times in both temporal and content dimensions.}
    \label{fig:A11}
\end{figure}


\begin{figure}[htbp]
    \centering
    \begin{subfigure}[b]{0.45\textwidth}
        \centering
        \includegraphics[width=\textwidth]{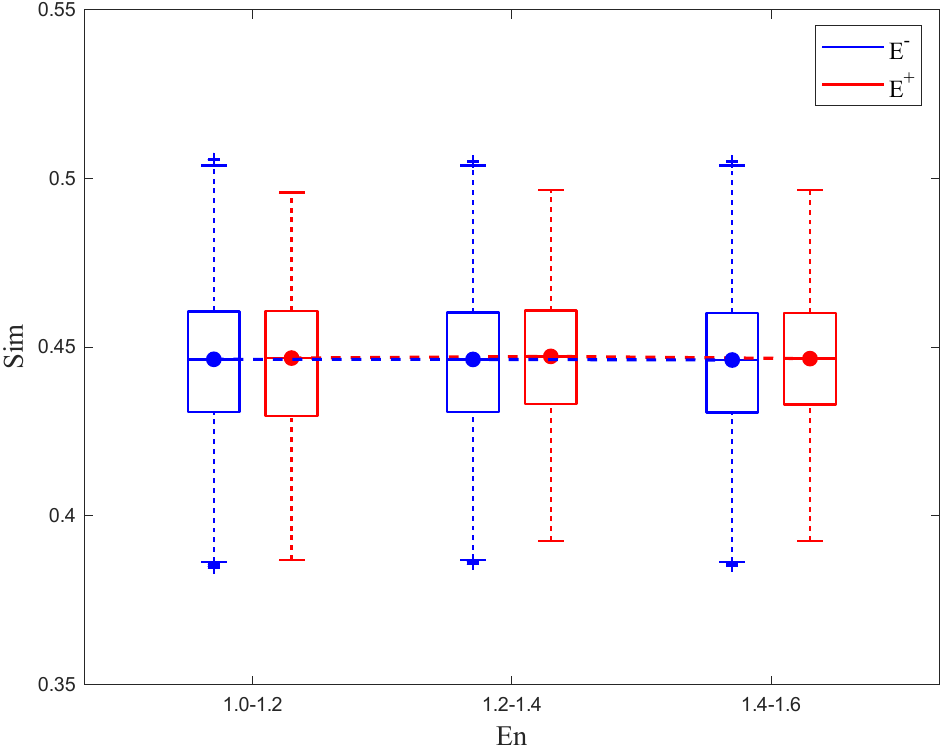}
        \caption{}
        \label{fig:A12a}
    \end{subfigure}
    \hfill
    \begin{subfigure}[b]{0.45\textwidth}
        \centering
        \includegraphics[width=\textwidth]{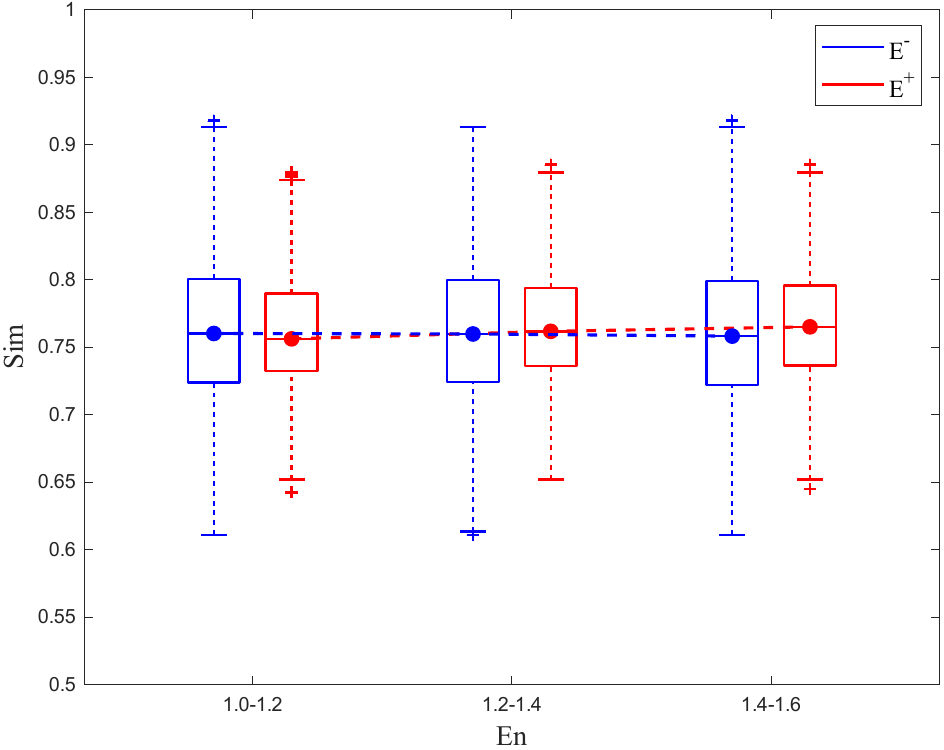}
        \caption{}
        \label{fig:A12b}
    \end{subfigure}

    \caption{The x-axis represents the entropy of the exposed content, and the y-axis denotes the similarity between each news item and historical content. The blue points ($E^-$) represent the news items that were exposed but not clicked, while the red points ($E^+$) represent the news items that were both exposed and clicked. The boxplot illustrates the distribution of the news similarity, where (a) denotes the median and (b) represents the maximum value. The difference between the blue and red points reflects the disparity between clicked and unclicked content.}
    \label{fig:A12}
\end{figure}

\end{document}